\documentclass[preprint]{aastex}
\begin{document}

 \title{A Photometric and Spectroscopic Study of the Cataclysmic Variable  ST LMi during 
 2005-2006\footnote{Based on observations from the WIYN Observatory, which is a joint facility 
 of the University of  Wisconsin-Madison,  Indiana University, Yale University, and the 
 National Optical Astronomy  Observatories,  USA.}, \footnote{Based on observations obtained 
 with the Mayall 4m telescope at Kitt Peak National Observatory, a division of the National 
 Optical Astronomy Observatory, which is operated by the Association of Universities for 
 Research in Astronomy Inc., under cooperative agreement with the National Science 
 Foundation}}

\author{S. Kafka\altaffilmark{1,2,3}}
\author{S.B. Howell\altaffilmark{3,4,5}}
\author{R.K. Honeycutt\altaffilmark{3,6,7}}
\and
\author{J.W. Robertson\altaffilmark{3,8,9}}

\altaffiltext{1}{CTIO/NOAO, Casilla 603, La Serena, Chile.}
\altaffiltext{2}{E-mail: skafka@noao.edu}
\altaffiltext{3}{Visiting Astronomer, Kitt Peak National Observatory, 
National Optical  Astronomy Observatory, which is operated by the Association 
of Universities for Research in Astronomy, Inc. (AURA) under cooperative 
agreement with the National Science Foundation.}
\altaffiltext{4}{WIYN Observatory \& NOAO, P.O. Box 26732, 950 N. Cherry Ave., 
Tucson, AZ 85719}
\altaffiltext{5}{E-mail: howell@noao.edu}
\altaffiltext{6}{Indiana University, Astronomy Department, 319 Swain Hall 
West, Bloomington, IN 47405}
\altaffiltext{7}{E-mail: honey@astro.indiana.edu}
\altaffiltext{8}{Arkansas Tech Univ., Dept. of Physical Sciences, 1701 N. Boulder ,
Russellville, AR 72801-2222}
\altaffiltext{9}{E-mail: jeff.robertson@atu.edu}

\begin{abstract}

We present orbit-resolved spectroscopic and photometric observations of the 
polar ST LMi during its recent low and high states. In the low state spectra, 
we report the presence of blue and red satellites to the H$\alpha$ emission line;
the velocities and visibility of the satellites vary with phase. This behavior 
is similar to emission line profile variations recently reported in the low 
state of AM Her, which were interpreted as being due to 
magnetically-confined gas motions in 
large loops near the secondary. Our low-state spectroscopy of ST LMi is 
discussed in terms of extreme chromospheric activity on the secondary star. 
Concurrent photometry indicates that occasional low-level accretion may be 
present, as well as cool regions on the secondary near L$_{1}$  
Furthermore, we report a new ``extreme low-state'' of the system at 
V$\sim$18.5~mag. 

Our orbital high-state spectroscopy reveals 
changes in the emission line profiles with orbital
phases that are similar to those reported by earlier high-state studies. The 
complicated emission line profiles generally consist of two main components. 
The first has radial velocity variations identical to that of the major 
emission H$\alpha$ component seen in the low state. The second is an additional 
red-shifted 
component appearing at the phases of maximum visibility of the accreting 
column of the white dwarf; it is interpreted as being due to infall 
velocities on the accreting magnetic pole of the white dwarf. At the 
opposite phases, an extended blue emission
wing appears on the emission line profiles.  We confirm the presence of a broad 
absorption feature near 6275$\AA$ which has been previously identified as Zeeman 
 $\sigma^{-}$ absorption component to H$\alpha$.  
This feature appears at just those phases when the 
accretion pole region is mostly directly visible and most nearly face-on 
to the observer.

\end{abstract}


\section{Introduction}

 Cataclysmic Variables (CVs) are short orbital period (hours to days) interacting 
 binaries containing a white dwarf (WD) primary and a K-M (or later) dwarf mass losing secondary 
 star. It is generally accepted that mass transfer occurs as the secondary overflows 
 its Roche lobe and matter falls toward the primary star through the inner Lagrangian point (L1). 
 The future of the accretion stream is then determined by the magnetic
 field of the WD: in magnetic systems the stream is channeled toward the magnetic poles 
 of the primary along its magnetic field lines; in non-magnetic systems a hot 
 accretion disk is formed around the primary, replenished by continuous mass transfer. 
 The abundance of CV properties  is reviewed 
 in Warner (1995) and will not be repeated here. Among the various subcategories, 
 there is one which includes both disk and magnetic CVs with large (up to 5 mag) apparently random drops in the optical  brightness; these are known as VY Scl stars. Although there is 
 no well-accepted explanation for the cause of these low states, it is clear that
  accretion from the secondary star is significantly diminished, often exposing the two 
 stellar components and
 providing a unique opportunity to study their photospheres.
 Sadly, low states in CVs are erratic in occurrence and duration, making observational study 
 a challenge. Therefore, we generally rely on photometric 
 monitoring efforts by professional or amateur astronomers in order to ascertain when a given system is in a low state.

Recently, low state observations of the magnetic CV prototype, AM Her, 
 revealed short outburst-like events in the optical light curves of the 
 system which were attributed to flares (activity) on the secondary star (Kafka et al.
 2005a). Later work (Kafka et al. 2005b) revealed components of the
 H$\alpha$ emission line arising from the vicinity of the secondary star. These features
 were interpreted as being due to stellar activity on the secondary and their study  provides a new 
 method to explore the extent and 
 variability of magnetic structures on the low-mass donor star.
 Stellar activity on the 
 secondary in CVs has been invoked
 to explain a plethora of phenomena, such as long-term light curve variations (Bianchini 
 1988) and high/low states (Livio and Pringle 1994). However, correlation of such phenomena  with stellar activity has yet to be fully explored. 
 In AM Her, prominence-like loops with gas velocities of $\sim$300 km/sec persisted for more 
 than three years (Kafka, Honeycutt \& Howell 2006), suggesting that they might be present even when the system is in a high state, albeit masked by accretion flux (Kafka et al. 2005b). 
 Recently the polar VV Pup exhibited similar structures in the  H$\alpha$ emission line during its low state (Mason et al. 2007). In VV Pup
 the H$\alpha$ emission structures were highly variable on timescales of several weeks.

 ST LMi (CW 1103+254) was introduced as a magnetic CV in the work of Stockman et al. 
 (1983), in which linear and circular polarization varied with the 114 min orbital
 period of the system.  Several high-state emission line components 
 (Stockman et al. 1983; Bailey et al. 1985) have been attributed to infall velocities and to
 features arising near the secondary star.  Most studies (Stockman et al. 1983; 
 Schmidt, Stockman \& Grandi 1983; Bailey et al. 1985; Cropper 1986; Cropper \& Horne 1994,
 Stockman \& Schmidt 1996) have concluded that the system is normally a one-pole accretor, in 
 which the active pole is self-eclipsed by the limb of the white dwarf for a major part of the 
 orbit. Peacock et al. (1992) reported polarimetric evidence for accretion onto
 a second pole at times.  Schmidt, Stockman and Grandi (1983) provided a 
 geometrical model of the system, according to which the secondary star slightly lags 
 the single accreting magnetic pole. Using near-IR data, Ferrario, Bailey \& 
 Wickramasinghe (1993) modeled the high state cyclotron emission from the active pole and derived a 
 magnetic field strength of 11.5$\pm$0.5MG.

 Since 1990, ST LMi has been a part of the long-term monitoring program of a 0.41-m 
 automated telescope located in
 central Indiana (RoboScope\footnote{Honeycutt~et~al.~1994 and refs
 therein}). The  1990-2004 light curve of the system, which appeared in 
 Kafka \& Honeycutt 2005 (KH05), revealed a ~5-year long low state (1992-1997), 
 followed by $\sim$6  years of high state (Fig~1). During the high state, the light curve of ST LMi 
 shows a 
 $\sim$1.5 mag orbital modulation caused by the self-eclipse of the single accreting pole by the 
 white dwarf (Stockman \& Schmidt 1996).  During the low states, the RoboScope light curve 
 showed that ST LMi usually reached a quiescent magnitude of 17.5 with an approximate sinusoidal 
 modulation of $\sim$0.2 mag (KH05).

 In this paper, we present a photometric and spectroscopic study of the system during 
 2005 and 2006. In Section 2 we describe our data and data reductions, in 
 Section 3 we discuss the results, while in Section 4 we compare these results to the
 behavior of other polars in the low state.  Our conclusions are summarized in Section 5.
 The results of this study show that the secondary star in ST LMi is a highly active star and behaves in a similar manner as those observed in a small but growing sample of  polar mass donors
 observed during low states. 
 
\section{Observations and Reductions}

\subsection{Photometry}

In the top panel of figure 1 we have combined our V-band photometry of ST LMi from 
various sources.  The data prior to 2004-Nov are RoboScope data from KH05.  RoboScope 
continued to collect photometry of ST LMi until 2005-Mar, which is
 also included in figure 1; after 2005-Mar RoboScope ceased operation (temporarily 
we hope).  The RoboScope data were processed/reduced through a pipeline that performs 
aperture photometry and extracts light curves of all the stars in the field based on 
the method of incomplete ensemble photometry (Honeycutt 1992).                   

The photometry later than 2005-Mar in figure 1 is mostly from the WIYN 
0.91-m telescope at Kitt Peak.  Processing of the 
non-RoboScope images followed standard IRAF\footnote{IRAF is distributed by 
the National Optical Astronomy Observatories, which are operated by the Association 
of Universities for Research in Astronomy, Inc., under cooperative agreement with the 
National Science Foundation.} procedures, using aperture photometry for the
extraction of instrumental magnitude. We used stars from Henden \& Honeycutt (1995) 
as secondary standards.  Much (but not all) of this 0.91-m photometry
consists of sequences of exposures covering $\sim$one orbital period; these sequences
are grouped into several observing runs, designated A-D (figure 1 and Table 1). Figures 2-4 show our phased photometric light curves of ST LMi, using only the data from figure 1
that is continuous for $\sim$1 orbit.  In this paper we have adopted an orbital ephemeris 
which uses the period from Cropper (1986) and a spectroscopic zeropoint (or epoch) from Howell et al. 
(2000).  The spectroscopic epoch is based on IR observation of photospheric absorption lines from the secondary star and phase zero marks 
the time of inferior conjunction of the secondary. For reference, phase zero for our ephemeris
corresponds to phase 0.17 on the polarimetric ephemeris of Cropper (1986), thereby placing the expected 
location of the observer's most direct view of the active accretion pole region near our phase 0.83. 
The A-D designations shown in fig. 1, and listed in Table 1, are used as appropriate in 
figures 2-3.  We show phased RoboScope light curves in fig. 4.

\subsection{Spectroscopy}

 Our spectroscopic data were obtained in four different observing runs
 during the interval 2005-Jun to 2006-May (indicated in figure~1, bottom).  During runs 2005-Jun (one night) 
 and 2006-Feb (two nights) the HYDRA multi-object fiber spectrograph (MOS) on the 
 WIYN\footnote{The WIYN Observatory is a joint facility of the University of 
 Wisconsin-Madison, Indiana University, Yale University, and the National Optical Astronomy
 Observatories, USA.} 3.5-m telescope was used.  At WIYN we employed the 600 line 
 mm$^{-1}$ grating in first order, blazed at 7500$\AA$; the spectral 
 coverage was $\sim$5500-8000$\AA$, with a resolution of $\sim$3$\AA$. A number of 
 non-assigned fibers were placed at random locations to serve for
 sky subtraction, and a CuAr lamp was used for wavelength calibration.
 The R-C Spectrograph on the 4-m KPNO telescope was used to acquire two spectra of 
 ST LMi in 2005-Dec, and to obtain a sequence of spectra covering a little more than one orbit
 in 2006-May. For the 2005-Dec spectra
 we used the KPC-007 grating and the GG-385 order separation filter. The spectral 
 coverage was $\sim$5500-7500$\AA$ with a resolution of $\sim$3$\AA$.   For the 
 2006-May data we used grating KPC-24 in second order with a 1" slit to match the seeing. 
 This combination provided $\sim$1000$\AA$ coverage at 1.3$\AA$ resolution. 
 The 2006-May night was clear with photometric conditions and stable
 seeing, allowing extraction of good quality relative photometric information from these spectra. 
 We produced a synthetic light curve using the fluxed spectra from this night of high state spectroscopy (figure 3 bottom right, see \S4.2). An instrumental magnitude was extracted from the reduced 4-m spectra
 using imstat to sum the counts in rectangular windows (including both lines and continuum) for star and sky.
The length of each window was $\sim$80$\%$ of the length of the spectrum,
yielding an effective wavelength approximately midway between V and R. Exposure 
 times for all our spectra ranged between 300 and 1800 sec, depending on the weather conditions (Table 1).   

Data processing and reduction employed standard 
IRAF procedures such as $\it{twodspec}$ with wavelength calibration accomplished via the use of
arc lamp exposures obtained near in time and position to our stellar exposures.
For the R-C spectrograph data, we extracted the 2-d images, reduced them in the usual fashion, flux calibrated them using the spectrophotometric standard star GD140, and produced final 1-d spectra. No corrections were made for 
atmospheric extinction as the slit was kept at the parallactic angle and the total range of airmass was small and never larger than 1.6. The usual telluric absorption features remain in all the final 
spectra. For Hydra, we performed the additional steps to allow correction for scattered light and the more complex sky subtraction necessary for a fiber spectrograph.  Partly cloudy weather conditions prevailed during all our runs with the exception of the 2006-May observations.


\section{Photometric Results}

  The photometric coverage in figure 1 is not continuous during 2005-2006 (due to the
  untimely failure of RoboScope) but the available photometry does show that 
  ST LMi was in a nearly-continuous low state
  during most of our photometric and spectroscopic observations.  In 2005-2006,  
  the most common basal magnitude in the low state appears to be $\sim$17.5 (bottom panel of figure 1); 
  consistent with the brightness during the 1992-1996 low state recorded by RoboScope
  (top panel of figure 1). However, the photometric sequence of 2006-Feb-12 (UT) (sequence C1 in Table 1)
  shows that the system reached V$\sim$18.0-18.4, significantly fainter than other data points
  in figure 1. The 0.91-m exposures showing this unusually faint state were acquired using 
  the same equipment, and were reduced using the same procedures as our other photometric observations near this time (see table 1) 
  which show the system at V$\sim$17.5.  We therefore conclude that this unusually low
  state is real.  Although it appears inconsistent 
  with the rest of the RoboScope data (which never showed such a state over 13 years), the 
  RoboScope magnitude limit is about 17.8; therefore fainter events would escape detection.
  Although unusual, the presence of an ``extreme-low state'' is not unique to ST LMi; 
  VV Pup, for example, was also observed to have two different low state levels (Mason et al. 2007). This extreme low state, allows for an interesting experiment for ST LMi:
assuming that the extreme low state (mean V=18.2mag) represents a time where
activity and accretion temporarily stop then the observed light in V should
be dominated by the white dwarf alone. For a DA WD, M$_{V}$=12.5. This gives a
distance for ST LMi of 138pc. This is very similar to the distance of 136 pc derived by
Bailey et al. (1985) using the K surface brightness for an M5 main sequence
star and a secondary mass of 0.18 M$_{sun}$.

 The low state orbital light curves from our photometric monitoring are 
 shown in figures 2 and 3. Light curves A1-A5 show a 0.3-0.8 mag ``bump'' near 
 phase 
 0.8, the phase at which we most directly view the main accretion
 region on the magnetic white dwarf.  Although the mean magnitude of these sequences
 was about at the 
 ``normal'' low state brightness, the observed bump in the light curves suggest that some 
 low-level variable 
 accretion (likely wind accretion) was taking place in 2005-late Feb/early March.  In contrast, the low state 
 orbital light curves for 2005-Jun and 2006-Feb (B1-B4, D1) appear consistent with 
 the 
 1992-1996 RoboScope low-state orbital light curve shown in fig.  4, bottom panel). These low state light curves are  essentially 
 featureless with 
 (at best) only a $\sim$0.1 mag  sinusoidal variation superposed on 
 $\sim$0.2 mag random variations.   Note that even the random 
 variation is largely missing for the low-state orbital light curve for 
 UT 2006-Feb-21 (fig. 3, D1).   
In 2005-March-1 (UT) we monitored ST LMi 
 (A2) using an R filter. Figure 3 
 (right, top) shows our phased differential R band light curve with an arbitrary zero point. 
 Again we see an orbital bump during this low state observation but with a larger amplitude than observed in V at nearly the same time (fig. 2). Even with a slightly variable amplitude from night to night, the larger amplitude bump observed in R compared with V during the low state agrees with previous high state observations of ST LMi showing that it becomes redder near phase 0.8.  
We will see below that for ST LMi, the red color at this phase is due to the contribution of the cyclotron continuum to the overall flux, a contribution that outshines the other components during this phase interval.

  It is of particular importance to ascertain the photometric state at the time of
  the spectroscopic observations, using both the brightness of the system and the orbital
  light curve.  For the 2005-Jun-17 WIYN spectra, simultaneous photometry 
  (figure 2, sequence B2)
  shows V$\sim$17.5 magnitude, with little or no orbital modulation.
  Therefore the system was in the normal low state, with no 
  evidence of accretion.  For the 2005-Dec-31 4-m spectra we have V-band measures on
  two nights, 7 and 8 nights preceding the spectroscopy, from Tenagra 
  Observatory\footnote{http://www.tenagraobservatories.com}, showing
  ST LMi at  V$\sim$17.3, again indicating a normal low state.  During the
  2006-Feb-21/22 (UT) WIYN spectroscopy, we have simultaneous WIYN 0.91-m photometry 
  showing the system at V$\sim$17.5.   The orbital light curve for 2006-Feb-22 (UT) 
  (sequence D1, figures 1 and 4) is very ``quiet'', with no significant orbital variations.
  Therefore the system showed no evidence of accretion for 2006-Feb-21 (UT)
  (based on its brightness) nor on 2006-Feb-22 (UT) (based on both the brightness and
  on the flat orbital light curve).  The 2006-Feb-21/22 data is the nearest photometric
  measurement to our 2006-May-19 4-m spectroscopy available to us.  
  In order to better determine the photometric state for the 2006-May spectroscopy
  we produced a relative light curve using the summed flux in the 4-m spectra (see section 2.2 for details). The transparency and seeing on this night were very good and
  stable, resulting in a quantitatively good light curve 
  (figure 3, bottom right) at an effective wavelength
  which is approximately mid-way between V and R, and has an arbitrary zeropoint.  This 
  light curve has a conspicuous 1.7 mag ``bump'' near phase 0.8, similar to that seen
  in high-state photometry such as the RoboScope high-state
  orbital light curve (top panel, figure 4\footnote{the phasing of the folded light curves 
  of ST LMi in KH05 (figure 5 in that paper) is in error because a wrong epoch for the 
  ephemeris was mistakenly employed.}). The bump at phase 0.8 is attributed to accretion onto the
  white dwarf pole as seen most directly by the observer. While some low state photometry shows
  residual accretion, likely due to wind accretion (e.g., sequences A1-A5, fig. 2) the amplitude and consistency of the bump seen in the 2006-May light curve (fig. 3) is similar to the RoboScope photometry high state light curve (fig. 4, top). Thus, it seems that during our 2006-May spectroscopy ST LMi was probably no longer in the low
  state; this will be discussed in Sec. 4.2.

 In summary, we conclude that the orbit-resolved WIYN spectroscopy of 2005-Jun and 2006-Feb 
 were in the low state, whereas the orbit-resolved 4-m spectra of 2006-May were 
 likely in the high state. 

\section{Spectroscopic Results}

\subsection{Low State Spectra}

Figure 5 shows representative spectra from the three low-state spectroscopic runs 
in 2005-Jun, 2005-Dec, and 2006-Feb.  All spectra were taken at a similar spectral resolution
of $\sim$3$\AA$ and clearly show the TiO bandheads near 7050$\AA$ from the secondary 
star, confirming our earlier conclusion that any accretion luminosity was very low at 
these epochs.  The H$\alpha$ emission 
line has considerable structure, as shown in the enlarged inset plots, similar to the 
profiles of the H$\alpha$ line of AM Her (Kafka et al. 2005b; 2006) and 
VV Pup (Mason et al. 2007) during their recent low states.

Figures 6 and 7 show the H$\alpha$ emission line profiles as a function of phase for two adjacent nights
of WIYN spectroscopy 2006-Feb-21/22 UT. In both nights, the emission line occasionally develops a red satellite near phases 0.75-1.00 and a blue component to the H$\alpha$ emission line for phases near 0.4 and 0.5. The complex emission line profiles have been fitted with Gaussian (in a 
manner identical to that described in Kafka et al. 2006 for AM Her) to determine their 
radial velocities (RVs).
We have been conservative in choosing between noise in the profile
and a real emission line component; we used satellites that exceeded the noise level by 2.5$\sigma$.  The RV curves for the line components we judged to be
reliable, are shown in figure 8.
Figure 8 also includes the RV 
 of the central peak of the H$\alpha$ line for the 2005-June (squares) and the 
2005-December (triangles) data. The RVs of the satellites are plotted with open symbols; 
different colors correspond to different cycles. The RV phasing of the central emission line
component of H$\alpha$ indicates that it arises from the secondary star's side of the
center of mass. The blue/red crossing is ~0.05 phase  units away from phase zero (which 
marks inferior conjunction of the secondary), but is consistent with the 
uncertainty of $\sim$0.1 phase units quoted by Howell et al. (2000) for the time of 
conjunction in the ephemeris. The velocities and phase visibility of the occasional 
blue and red satellites in figure 8 are quite similar to that those reported for 
AM Her (Kafka et al. 2005, 2006) and VV Pup (Mason et al. 2007); they have been interpreted
as arising from gas motions along two large magnetically-confined loops, perhaps
similar to slingshot prominences. 

 Figure 9 (top) shows the low-state equivalent width of the full H$\alpha$ emission line for the WIYN 
 spectroscopy of 2006-02-21/22 vs orbital 
 phase. For comparison, we reproduce the phased light curve (D1) of figure 3 on a 
 expanded magnitude scale so as to see the structure of the lightcurve. The H$\alpha$ emission line 
 appears to be stronger between phases 0.1-0.5 (albeit with variable strength); this  
 is the trailing side of the secondary star. Near the same phases 
 the light curve appears to have a V$\sim$0.05~mag dip, which is repeated in 
 two consecutive orbits. Since accretion was absent at this epoch, the light curve dip could be due to 
 the presence of a dark starspot region on the secondary star, with the 
 H$\alpha$ brightening corresponding to associated chromospheric activity. This behavior suggests that magnetic activity (spots) near the 
 L1 point may indeed be connected to the occurrence of the low states, in agreement with the Livio \& Pringle (1994) work.

\subsection{High State Spectra}

In spite of the lack of concurrent photometry we have concluded that ST LMi 
was likely accreting in the high state during our 2006-May 4-m spectra.
ST LMi is often considered a ``simple'' one-pole accretor, rather similar to VV Pup.
There is general agreement in the literature regarding the geometry of the system: 
the orbital inclination is near 55$\arcdeg$ and the main accreting pole has an 
inclination of $\sim$150$\arcdeg$ with respect to the white dwarf spin axis (e.g. 
Schmidt, Stockman \& Grandi 1983; Cropper 1986; Peacock et al. 1992;  Ferrario et al. 
1993).  This geometry leads to a self-eclipse of the main accretion region by the
white dwarf, providing good leverage for extracting geometrical properties of this
pole. The main accreting pole is the one nearest to the secondary star, having a magnetic field 
strength of $\sim$13MG). 
The derived accretion footprint leads (in the direction of orbital motion)
a line joining the stellar centers by $\sim$0.05-0.15 phase units (Schmidt et al. 1983; 
Cropper 1988; Howell et al. 2000) being most directly viewed at our phase 0.8. The main accretion region has been shown to have a 
complex extended pattern on the white dwarf (e.g. Cropper \& Horne 1994; Stockman \& 
Schmidt 1996).  The second pole has a strength of $\sim$30 MG and  has been reported 
to also accrete on occasion (Peacock et al. 1992). 

Figure 10 presents a representative spectrum from the 2006-May-19 4-m run.  H$\alpha$ is 
strongly in emission with 
a complex and variable profile. Furthermore the HeI emission lines at 
5876$\AA$ and 6678$\AA$ are quite strong, consistent with previous existing high state 
spectroscopy of the system, and inconsistent with the low state WIYN spectra in which the 
HeI emission lines are completely missing.

The phased high-state optical light curve of ST LMi (Stockman et al. 1983; Cropper 1986; 
Peacock et al. 1992; also see figure 4) has a bright phase lasting about 0.25 phase units 
centered near orbital phase 0.8, in which the system  becomes $\sim$1.2 mag brighter in the 
V-band.  The amplitude of this feature becomes larger as one moves from
B to V to R, consistent with our measurements. The bright phase corresponds 
to the time when the primary accreting pole is most directly viewed from the earth. The 
spectrum of ST LMi is 
significantly different between the bright orbital phases and 
the faint orbital phases of the high-state (see below).
The high-state optical emission lines are quite complex and have varying interpretations
 involving infall velocities
from an accretion funnel plus other components thought to be produced near the secondary star.
(Schmidt, Stockman \& Grandi 1983; Bailey et al. 1985).
This picture of the origin of the H and He emission lines during the high state bright phase is similar to that developed for other single pole accretors, such as VV Pup. In ST LMi, though, Schmidt et al. (1983) and Bailey et al. (1985) agree, as we discuss below, that some of the emission line components are from the secondary star and/or its vicinity. 

Stockman et al. (1983) identified two major components in their ST LMi high-state optical  
emission lines (H$\delta$, H$\gamma$, HeI4471, HeII 4686 and H$\beta$):
a broad emission component (BEC) which has a sinusoidal velocity curve with
K$\sim$500 km s$^{-1}$ plus a non-sinusoidal narrow emission component (NEC) which is 
always redshifted with respect to the BEC.  They conclude that the BEC emission arises 
near the primary star, in a cone-like structure having an infall velocity of $\sim$1000 
km/sec. The high velocities imply that the emitting region should be close to the white 
dwarf, in a curved, organized path that leads to the observed line broadening. Using the 
same data,  Schmidt et al. (1983) constructed a physical model 
for the accretion stream, and identified the BEC with a ``constricted region'' 
approximately mid-way between the two stellar components, where the accretion stream 
encounters strong magnetic pressure as well as becoming confined by the magnetic field of the white dwarf. 
The bright phase (0.5-0.95) NEC is attributed to a highly collimated infall  
of high-density gas, similar to the situation  observed in EF Eri in its high state (Schneider 
\& Young 1980; Young et al. 1982).  During the faint optical phase (0.0-0.5), the NEC 
(which dominates the peak of the emission) is thought to arise from the secondary star 
itself - possibly irradiation on the heated inner hemisphere. However, upon calculating the 
area that gives rise to this emission, these authors find that it is one order of magnitude less 
than the area of the irradiated inner hemisphere of the secondary star. This discrepancy 
is interpreted as being due to limb-darkening effects which were not taken into account in 
the calculation, and to a degree of self-eclipsing by the accretion funnel.

The high state data of Bailey et al. (1985) found structure in  
the HeII 4686$\AA$ and H$\beta$ emission lines similar to that reported in the Stockman et al.
(1983) study. However, these authors argue that the profiles are made up of two
sinusoids, one of which has variable width with phase. Component A appears narrow and 
reshifted for the 
bright part of the orbit, reaching velocities of 600 km/sec, whereas it is faint and broad
at other phases.  As in the work of Stockman et al. (1983) and Schmidt et al. (1983)
this component is attributed to a highly curved inflow
region onto the white dwarf. The second component (component B) has a sinusoidal velocity 
curve with K$\sim$300 km s$^{-1}$ and was attributed to the secondary star - either its 
irradiated inner hemisphere or material in its vicinity.

Our phase resolved profiles of the H$\alpha$, HeI 5876$\AA$ and HeI 6678$\AA$ emission lines (figure 11) are complex, having many of the same features as shown in the earlier
high-state studies of ST LMi.
In particular, we see double-peaked emission lines
between phases 0.6 and 0.9, plus an extended blue wing for phases 0.1-0.4. 
The profiles are difficult to fit with multiple Gaussians because of their
asymmetric nature plus the uncertainty regarding the number of components.  For those
reasons we chose to characterize the radial velocities using two simple methods:
1) the centroid of the full emission line, regardless of the structure, and 2) the 
velocities at the emission line peaks of one, or at most two, well-defined components.  
This is similar to the approach of Bailey et al. (1983), whereas Stockman
et al. (1983) measured velocities of emission line components using bisectors and centroids.  
We find no significant differences in the shapes or velocities of the He and H$\alpha$
emission lines. 

In figure 12 (top) we show the measured radial velocities of the 
centroid of the full H$\alpha$ emission line (open triangles) along with the velocities of the 
individual peaks.
For comparison, we reproduce the velocity  curve of the central H$\alpha$ component from the 
low-state WIYN data (crosses). The other two panels include the measured RV of the centroid of the full HeI 5876$\AA$ (middle) and HeI 6678$\AA$ (bottom). In the top panel of fig. 12 we see that, for about half the orbit (phases 0.0-0.6), the centroid of the
full H$\alpha$ emission line (open triangles) follows the curve established by the 
2006-Feb-21/22 low state data (crosses), as does the radial velocity of the single peak of 
the line (solid points) over phases 0.0-0.6.  However, over the
phase interval 0.6-1.0 the centroid deviates sharply redward as do the two distinct emission line 
components  (solid points).  
The redder component eventually deviates
from the low-state curve by nearly 800 km s$^{-1}$.  The radial velocity of the peak
of the blue component also deviates somewhat redward from the low-state curve.  However, this 
is likely due to contribution of underlying emission from the redder
component.  We therefore will proceed under the assumption that the
major change that occurs over phases 0.6-1.0 is the appearance of a new emission line
component that is redshifted 700-800 km s$^{-1}$ with respect to the persistant component.

Ferrario \& Wehrse (1999) have computed model line profiles for polars,
finding a narrow, low-velocity component from the magnetically heated
coupling region, while a second component arises from inflow near the
white dwarf.  We think that the line component we observe over phases 0.6-1.0
is indeed due to inflow near the white dwarf, but identifying the "other"
line component in our spectra as being from the coupling region does not seem
to be possible.  The magnetic
field in ST LMi is similar to that in EF Eri, for which the distance of
the coupling region above the white dwarf is estimated to be 5R$_{WD}$
(Meggitt \& Wickramasinghe 1989).  If the coupling region in ST LMi is
also $\sim$5R$_{WD}$, we find this emission to arise $\sim$0.13a above
the white dwarf, while we find (using the system parameters from Stockman \&
Schmnidt (1996) that the center of mass is $\sim$0.29a above the white dwarf. 
This is well on the
white dwarf's side of the center of mass whereas the phasing of the 
"other" component in our spectra places the origin well on the secondary's
side of the center of mass.

In figures~13 through ~16 we present our high-state spectroscopy of ST LMi from 2006 May.
The spectra are shown with assigned phases and each panel has the same identical flux scale. 
The spectra at phases 0.917, 0.764, and 0.865 correspond  to the time of the orbital photometric bump, that is, the time when
the accretion pole region is most nearly face-on to the observer (see figure 4 top, for comparison).
These spectra have two features not seen at other orbital phases:
1) they have an absorption
feature near 6275$\AA$ and 2) they have a broad continuum hump from $\sim$6400$\AA$ redward. 

The broad absorption feature near 6275$\AA$ was also reported by Schmidt et al. (1983).
Those authors considered several possibilities for the identification: $\lambda$6280 TiO
from the secondary star, the H$\alpha$ $\sigma^{-}$ Zeeman absorption component from
 the white dwarf photosphere, and the 
telluric a-band (due to O$_{2}$).  Their preferred identification was the H$\alpha$ $\sigma^{-}$
line component from the white dwarf photosphere. Our improved spectral resolution
allowed us to make a more careful comparison with the a-band.  We used the very high
resolution telluric spectrum above Kitt Peak from Hinkle, Wallace \& Livingston (2003),
convolved with a Gaussian to match our spectral resolution.  The match
with the a-band is surprisingly good in wavelength, in approximate shape (degraded
to the red in the average spectrum of phases 0.917, 0.764, and 0.865), and in 
approximate strength. Unlike water vapor however, oxygen
is uniformly distributed in the earth's atmosphere and could not be modulated
with phase to appear only over this small orbital phase range, even by coincidence. 
Like Schmidt et al. (1983), we too 
can rule out TiO for a similar reason: the 6275$\AA$ feature appears only during a small range in orbital phase. Both of these features occur only during the time when the observer has the most direct view of the white dwarf magnetic pole. 
Given the above,  we  also conclude that the 6275$\AA$ absorption feature is due to the  H$\alpha$ Zeeman $\sigma^{-}$ line.

Examination of averaged WIYN low state spectra (figure 17) reveals the same Zeeman absorption component from H$\alpha$ as we discussed above. The wavelength of the $\sigma^{-}$ component can be used to calculate the surface magnetic field strength of  the WD. Note, this field
strength will be different from that estimated from the cyclotron humps as it comes from the 
surface while the cyclotron emission comes from somewhere in the column above the magnetic pole, probably, in fact a small range of locations. The separation of the Zeeman $\sigma^{-}$ absorption component from the central $\pi$ component of H$\alpha$ is 340$\AA$. Eq. 10 in  Wickramasinghe \& Ferrario (2000) yields a B of 16.8 MG for this Zeeman separation in H$\alpha$.

Looking at fig. 15 we see a redward hump in the continuum for those data obtained during the phase interval 0.7-0.9. The location of the hump corresponds to the n=14 harmonic for a B=12~MG field. This harmonic would be weak, if present at all. Cyclotron humps are far more prominent during times of very low mass accretion (e.g., Schmidt et al. 2005).
We therefore argue that this redward rise in the continuum slope is not a cyclotron hump but the general cyclotron continuum (the Wien slope) observed during this time of high state mass accretion and seen during the interval of near face-on viewing of the accretion region.

The geometry of ST LMi's orbit as well as main accreting pole location allows a unique view of a magnetic CV. The accretion pole is not directly visible for long (or at all) and the binary inclination provides both a near-direct view as well as places the observer in a near perpendicular plane to a normal to the field lines at the pole. Thus, we can observe Zeeman absorption from the near pole white dwarf surface as well as the beamed cyclotron emission from the accretion column simultaneously. The fact that we do not see a direct pole or only for a small time period (phases 0.75-0.85), allows us to study the
component stars during both high and low state with little contamination form the accretion flux.
 
\subsubsection{Hyperactivity on CV secondary stars}

Magnetic activity on the mass-losing secondary star of CVs has been previously inferred from
 a variety of types of excess emission. Low state x-rays have been attributed to coronal 
 emission from the secondary star in AM Her (de Martino et al. 1998), and  
 ST LMi is one of a few systems for which chromospheric activity has been deduced from 
 structures in the H$\alpha$ emission line of low-state optical spectra. There are currently only a 
 handful of  CVs whose low-state H$\alpha$ profiles suggest chromospheric activity (hyperactivity). Apart 
 from AM Her (Kafka et al. 2005b, 2006), VV Pup (Mason et al. 2007) and ST LMi, AR UMa 
 (Schmidt et al. 1999) and TT Ari (Shafter et al. 1985) have been reported to show extra 
 components in their low-state H$\alpha$ emission line. Although these 
 extra components in AR UMa and TT Ari have not been attributed to prominences
 on the secondary star, the emission line profiles and RV behavior are similar to the ones reported for AM Her and ST LMi. In all cases, the blue/red satellites are present for a portion of
 the orbit and, with the exception of ST LMi, the red satellite is present during the 
 first part of the orbital cycle and the blue during the second part.  In some instances both 
 satellites are present at the same phases (triple-peaked emission lines), and  
 in all cases there is a rapid switchover from red to blue satellite near spectroscopic 
 conjunction. The persistence/duration of the satellites varies from system to system: 
 in AM Her the satellites appear to have remained relatively unchanged for 3 years, 
 the satellites in VV Pup appeared only during a second epoch of spectroscopic observations 
 of the system and in ST LMi the visibility of the blue/red H$\alpha$ satellites changes 
 from cycle to cycle. Differences in the character and duration of activity is expected 
 for different spectral types of secondaries and different rotation rates 
 (i.e. system orbital periods). 

It is also worth noting that the systems with reported hyperactivity are polars 
(diskless MCVs), with the exception of TT Ari. Although
 TT Ari is classified as a nova-like CV, Jameson et al. (1982) proposed that the system 
 likely has a magnetic white dwarf therefore it belongs in the DQ Her category. An upper limit 
 of 4MG (Shafter et al. 1985) for the magnetic field of the primary is still within the range 
 of the magnetic field strength of the WDs in DQ Her (intermediate polar) systems. The suspected magnetic nature 
 of TT Ari, along with the known magnetic nature of the other systems, suggests that a 
 combination of fast rotation (all systems have orbital periods less than 4.0 hours) and 
 magnetic field interaction between the two stellar components may be responsible for the 
 presence of the satellites. Activity-induced line emission appears to be concentrated around 
 the inner Lagrangian point, although the observed velocities suggest that the relevant 
 structures are not confined to the orbital plane of the system. Again, an exception to 
 this rule is presented by TT Ari, in which the H$\alpha$ emission line appears to be stronger at 
 inferior conjunction of the secondary star.

 Triple-peaked H-alpha emission lines are not the only characteristic of stellar activity in CV 
 secondaries: EF Eri, after spending seven years in a low state, and without 
 changing its optical brightness, began showing weak and narrow Balmer lines 
 and CaII H \& K emission, originating on or near the secondary star (Howell et al. 
 2006). Although there are indications that the EF Eri H$\alpha$ emission line shows additional 
 components, 
 the S/N of the data was not adequate to reach secure conclusions about the structure of the 
emission line.  On the other hand low states in 
 disk systems are more rare than low states in MCVs.  Chromospheric activity and 
 activity cycles have been considered 
 responsible for quasi-periodic modulations in the long-term optical light 
 curves of dwarf nova in quiescence. Furthermore, variations (positive and negative) 
 in the timings of eclipse minima in disk systems have been explained in terms of magnetic 
 activity on the secondary (Baptista et al. 2003). Perhaps the most ``direct'' 
 observation of stellar 
 activity in a disk CV is provided by Steeghs et al. (1996) who described a peculiar 
 stationary 
 low velocity component in several emission lines of the quiescence spectra of the dwarf
 novae IP Peg and SS  Cyg. The authors attributed this component to slingshot prominences 
 (Collier Cameron \& Robinson 1989), co-rotating with 
 the secondary star and kept in the vicinity of the L1 point by the binary potential. Taken 
 at face value, this implies that the secondary stars of $\it{all}$ CVs should be 
 chromospherically active. Since the work of Parker (1961), it is known that 
 a tachocline (convective zone - radiative core boundary) is necessary for the formation 
 of a large scale, $\alpha\Omega$ (dipole) field, giving rise to all our familiar 
 (from the sun) activity phenomena. Rotation can amplify the field giving 
 rise to large-scale magnetic structures. Furthermore, magnetic field interactions may increase the lifetime of such structures. So far, the systems that show hyperactivity have secondary stars of spectral type M3 or later. Near that spectral
type, main sequence stars become fully convective likely leading to a change in the nature of
their magnetic field structure, because the formation of an $\alpha$$\Omega$-induced
 magnetic field is no longer possible.  For fully convective stars, chromospheric 
 emission may be generated by a small scale, local
turbulent magnetic field (Durney et al. 1993). Although this might be an alternative to   
the  $\alpha$$\Omega$ dynamo mechanism, it is not expected to produce large-scale
features such as large starspots, or a dominant dipole field for magnetic braking.
Other mechanisms might allow a fully convective star to produce a large scale field, such
as an $\alpha^{2}$ dynamo (Chabrier and Kuker 2006).  Fast rotation can amplify this field, 
which can easily reach the order of several kG, similar to the one that is 
observed in single active late M dwarfs, and similar to the one necessary for the 
observed activity on CV secondaries.

The He I emission line ratio is often used as an indicator of the physical conditions 
in the chromospheres of solar-type stars.  It can be shown (Howell et al. 2006) that
the ratio, EW[HeI(5876$\AA$)]/EW[HeI(6678$\AA)$] can be as high as 45 in the quiet sun but is near 
3 in active prominences. Using our 4-m high-state ST LMi spectra, we find that
the ratio of EW[HeI(5876$\AA$)]/EW[HeI(6678$\AA$)] emission lines is $\sim$2.5-3.0, with the exception of 
phases 0.7-0.9 when the 
accretion pole likely contributes to the line emission.  
This line ratio value is that expected for He I emission caused by collisional
excitation in an active chromosphere hotter than 8000K.

\section{Final Remarks}

 We have presented a spectroscopic and photometric study of ST LMi during its 2005-2006 high and low states. Our main results can be summarized as follows:

\begin{itemize}

\item In the low state, ST LMi remains at V$\sim$17.5~mag,
with an occasional "hump" in the optical light curve, between phases 0.5 and
0.95. This is the same phase interval during which the ``bright'' phase in
the  high state occurs due to the visibility of the accretion pole. Therefore
episodic low-level accretion apparently takes place in ST LMi during the low
state.

\item The low-state spectra of the system are dominated by strong TiO bands from
 the secondary star. The only emission line is H$\alpha$ and it appears to have structure
 similar that of AM Her (Kafka et al. 2005b, 2006) and VV Pup (Mason et al. 2007)
 in their low states. At times
 when accretion seems to have ceased completely (according to the optical light curves),
 the RVs of all components of H$\alpha$ originate from the
  secondary's side of the center of mass. Concurrent photometry supports the lack of accretion and the presence of a large cool region (spot) near the L1 point (between phases 0.3-0.6).

\item  A photometric state $\sim$ 0.7 mag fainter than the usual low state magnitude
was observed.  This is the second CV which
displays such a state in its optical light curve. It is not clear whether this was
a more complete cessation of mass transfer, or some other change in the system.
It will be important to study such events if we to understand the properties of
a CV in which accretion has truly stopped.

\item In the high state, the H$\alpha$ line presents a complicated multi-component profile.
 For orbital phases between 0.0 and 0.5, the RV of the H$\alpha$ emission line nearly coincides
 with the  RVs of the low state H$\alpha$ emission.  This line component, which is
 phased with the motion of the secondary star, cannot be associated with accretion
 because it is present in both the high and low state.  It is apparently not due
 to irradiation of the inner hemisphere of the secondary star because it is not
 brightest when the inner hemisphere most nearly faces the observer. It may 
 be due to activity on the secondary star, although the geometry of the emission region
 is far from clear.
 
\item In the high state a second emission line component of H$\alpha$ appears near phase 0.8 
which is highly red-shifted, likely due to infall onto the white dwarf.  We confirm the
results of Bailey et al. (1983) that a broad blue wing appears on the H$\alpha$
profile at phases opposite the appearance of the highly red-shifted component.

\end{itemize}

Although CV secondaries are 
the fastest rotating lower main sequence stars known, the large-scale activity signatures 
we have deduced are not expected for a fully 
convective star. It will be necessary to obtain suitable spectroscopy of a non-magnetic
(disk) CV in the low state to test if the presence of H$\alpha$ satellites 
are tied to magnetic field interactions of the two stars, or if ultra-fast rotation along suffice for their presence.

\acknowledgments
 We would like to thank our anonymous referee for the careful review of the manuscript.

\pagebreak  

\begin{deluxetable}{ccccc}
\tablenum{1}
\label{t:ages}
\centering
\tablecaption{Log of Observations of ST LMi}
\tablewidth{0pt}
\tablehead{ \colhead {UT Date}
& \colhead {Seq. No.}
& \colhead {Telescope/Instrument}
& \colhead{No. of Exposures/Filter}
& \colhead{Exp time(s)} }
\startdata
PHOTOMETRY:        &    &                 &        &     \\
1991-01 to 2005-03 &    & RoboScope 0.4-m & 551/V  & 240 \\
2005-02-28         & A1 & WIYN 0.9-m/S2KB & 44/V   & 120 \\
2005-03-01         & A2 & WIYN 0.9-m/S2KB & 35/R   & 90  \\
2005-03-02         & A3 & WIYN 0.9-m/S2KB & 42/V   & 90  \\
2005-03-03         & A4 & WIYN 0.9-m/S2KB & 50/V   & 90  \\
2005-03-04         & A5 & WIYN 0.9-m/S2KB & 33/V   & 90  \\
2005-06-13         & B1 & WIYN 0.9-m/S2KB & 11/V   & 180 \\
2005-06-17         & B2 & WIYN 0.9-m/S2KB & 12/V   & 120 \\
2005-06-20         & B3 & WIYN 0.9-m/S2KB & 13/V   & 120 \\
2005-12-23         &    & Tenagra 0.8-m   & 1/V    & 200 \\
2005-12-24         &    & Tenagra 0.8-m   & 1/V    & 200 \\
2006-02-12         & C1 & WIYN 0.9-m/S2KB & 23/V   & 120 \\
2006-02-21         &    & WIYN 0.9-m/S2KB & 6/V    & 180 \\
2006-02-22         & D1 & WIYN 0.9-m/S2KB & 68/V   & 180 \\
SPECTROSCOPY: & &                   &    &            \\
2005-06-17    & & WIYN 3.5-m/Hydra  & 6  & 600 - 1800 \\
2005-12-31    & & KPNO 4-m/RC Spec  & 2  & 1200       \\
2006-02-21    & & WIYN 3.5-m/Hydra  & 12 & 600        \\
2006-02-22    & & WIYN 3.5-m/Hydra  & 31 & 600        \\
2006-05-19    & & KPNO 4-m/RC Spec  & 16 & 300-600    \\

\enddata
\end{deluxetable}


\begin{figure}  
\epsscale{1.0}
\plotone{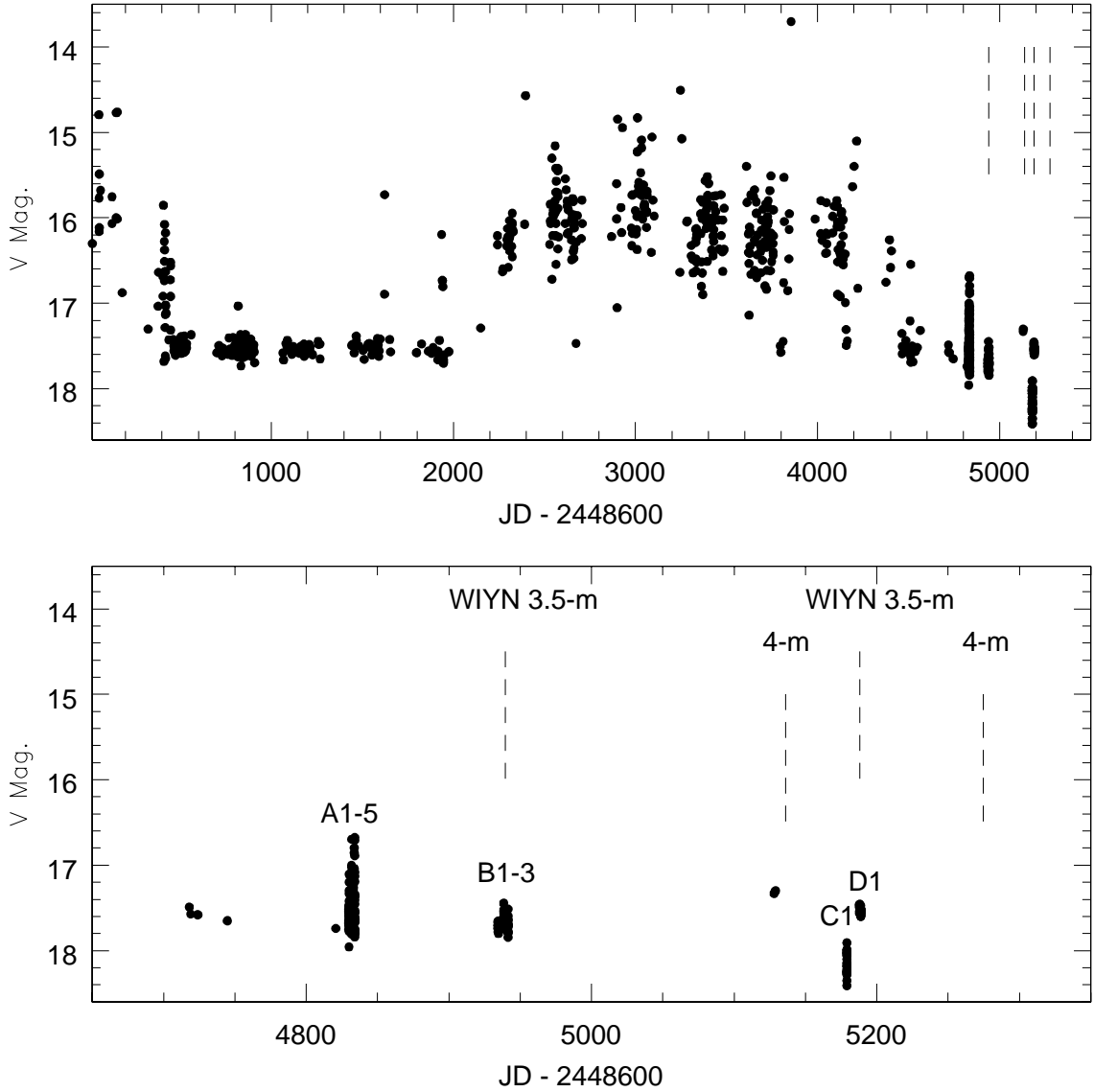}
\caption{Top: V-band light curve of ST LMi 1990-2005 from various sources, mostly
RoboScope.  The dates of our spectroscopy are noted as vertical dashed
lines.  Bottom:  Expanded view of the latter portion of the top panel, with the
spectroscopic observing runs noted.  The labeled photometric sequences correspond to the designations in figure 2 and Table 1.}
\end{figure}

\begin{figure}  
\epsscale{1.1}
\plottwo{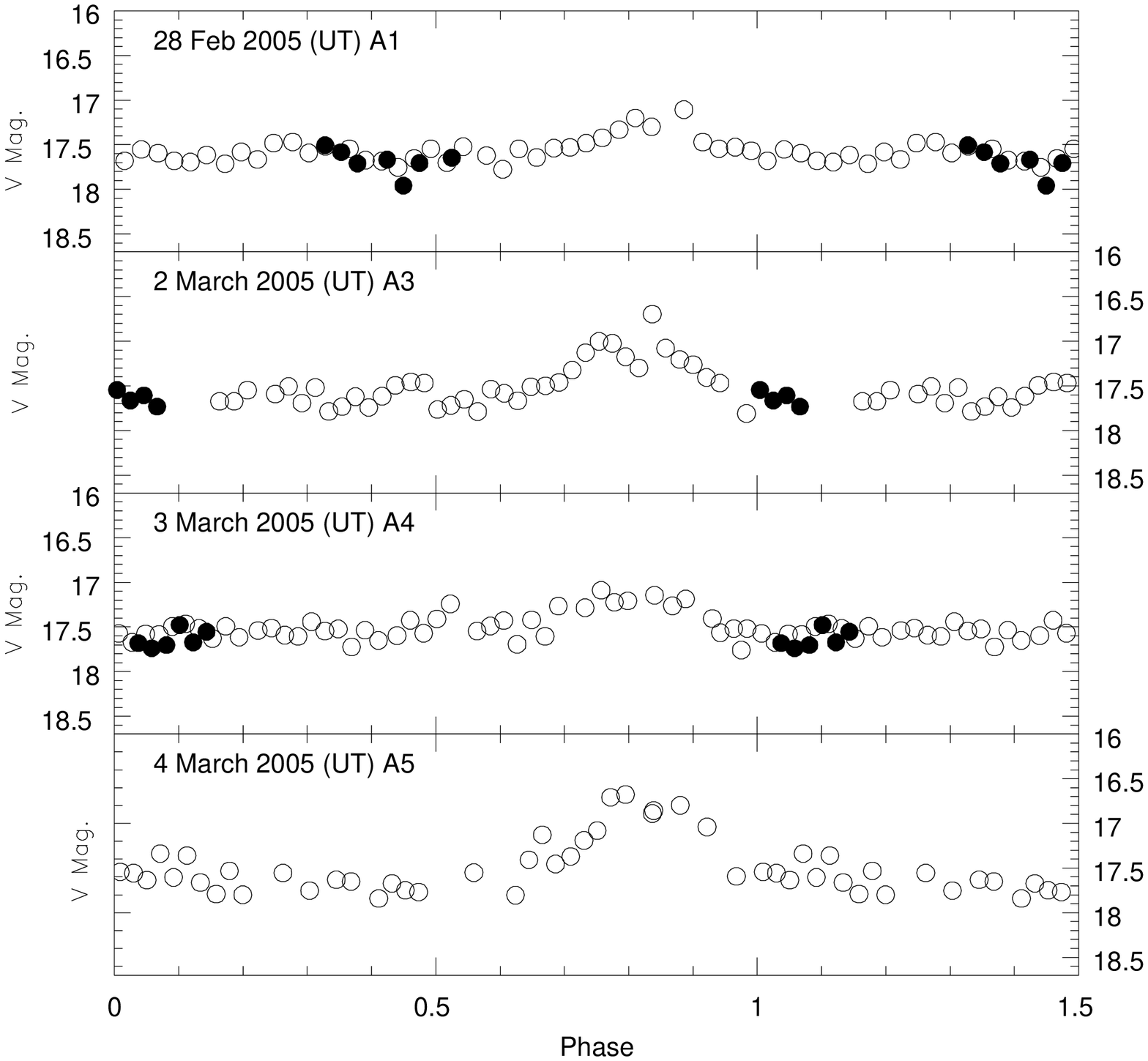}{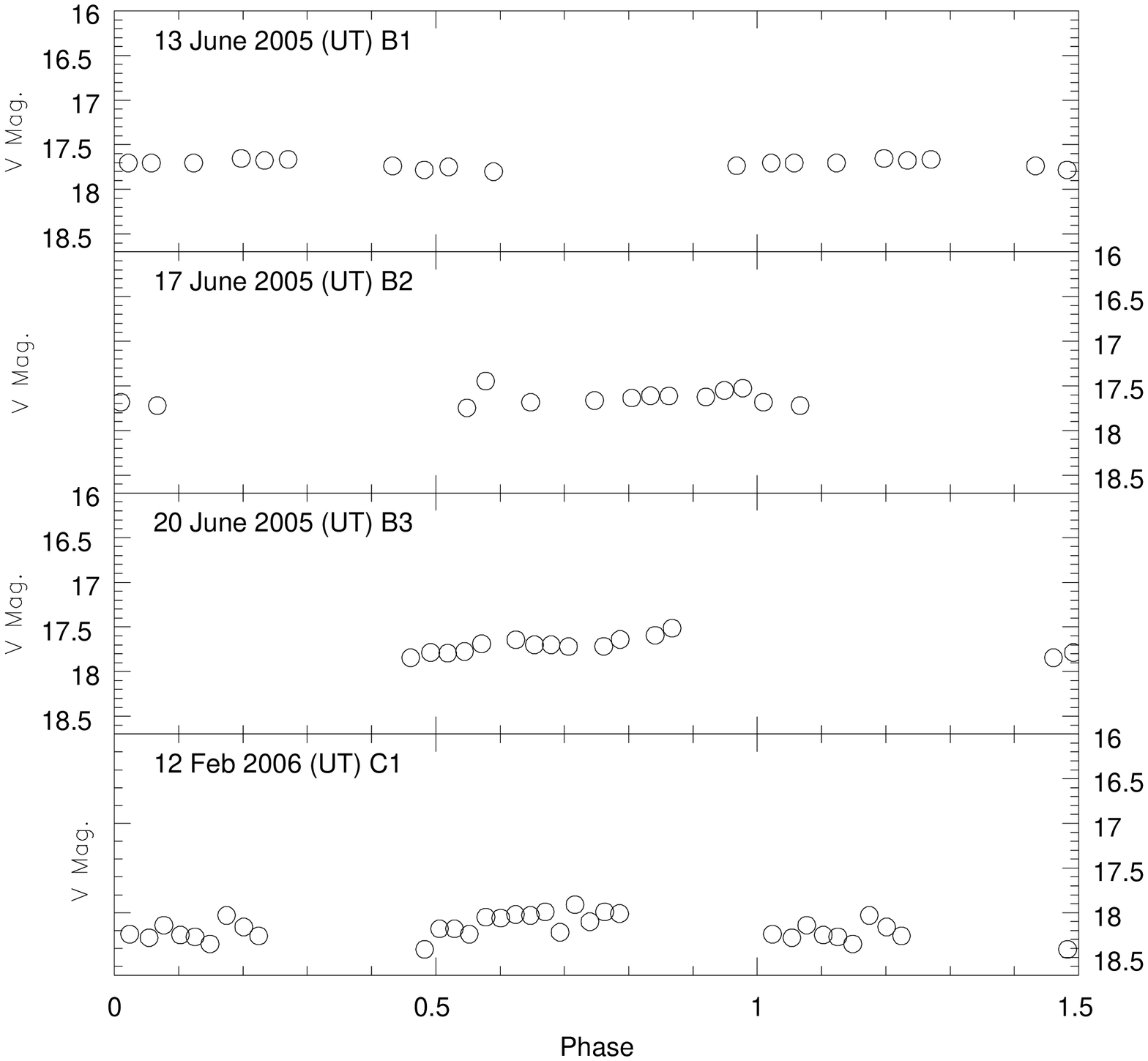}
\caption{Left: Phased V-band light curves of ST LMi from 4 nights of observing run A,
as marked in figure 1 (bottom).  Note the maxima near phase 0.8. Filled symbols
are data from the second of two adjacent orbital cycles.  Right: Phased V-band 
light curves of ST LMi from 3 nights of observing run B and
1 night of observing run C, as marked in figure 1 (bottom).  Note the absence of a
maxima near phase 0.8 }
\end{figure}

\begin{figure}  
\epsscale{1.1}
\plottwo{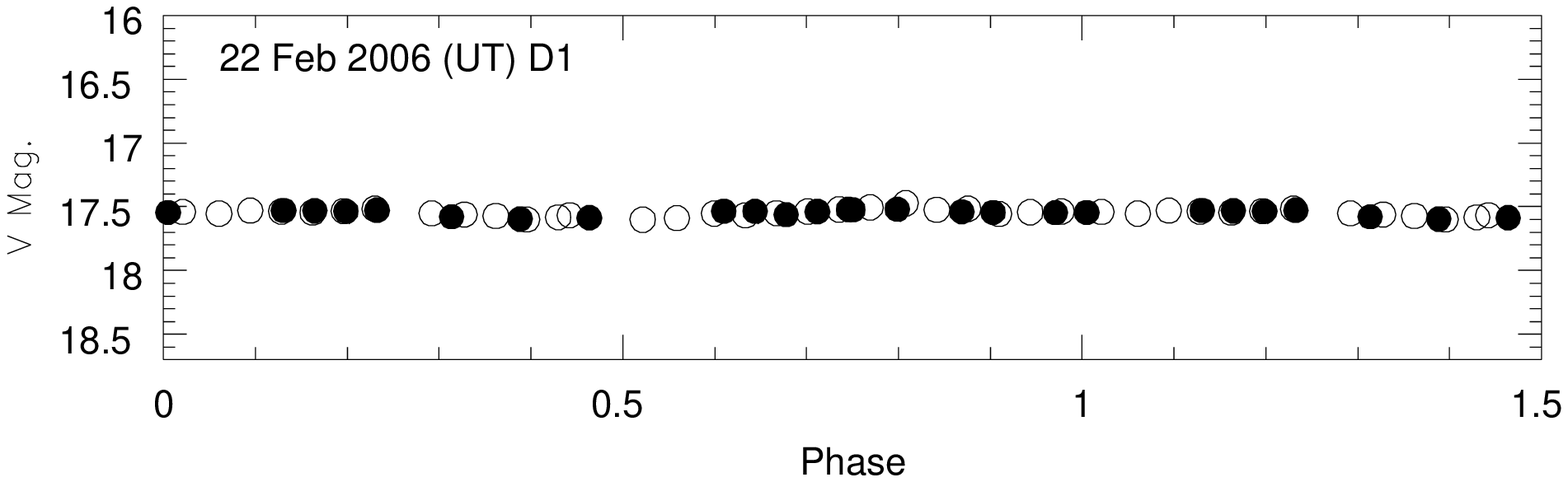}{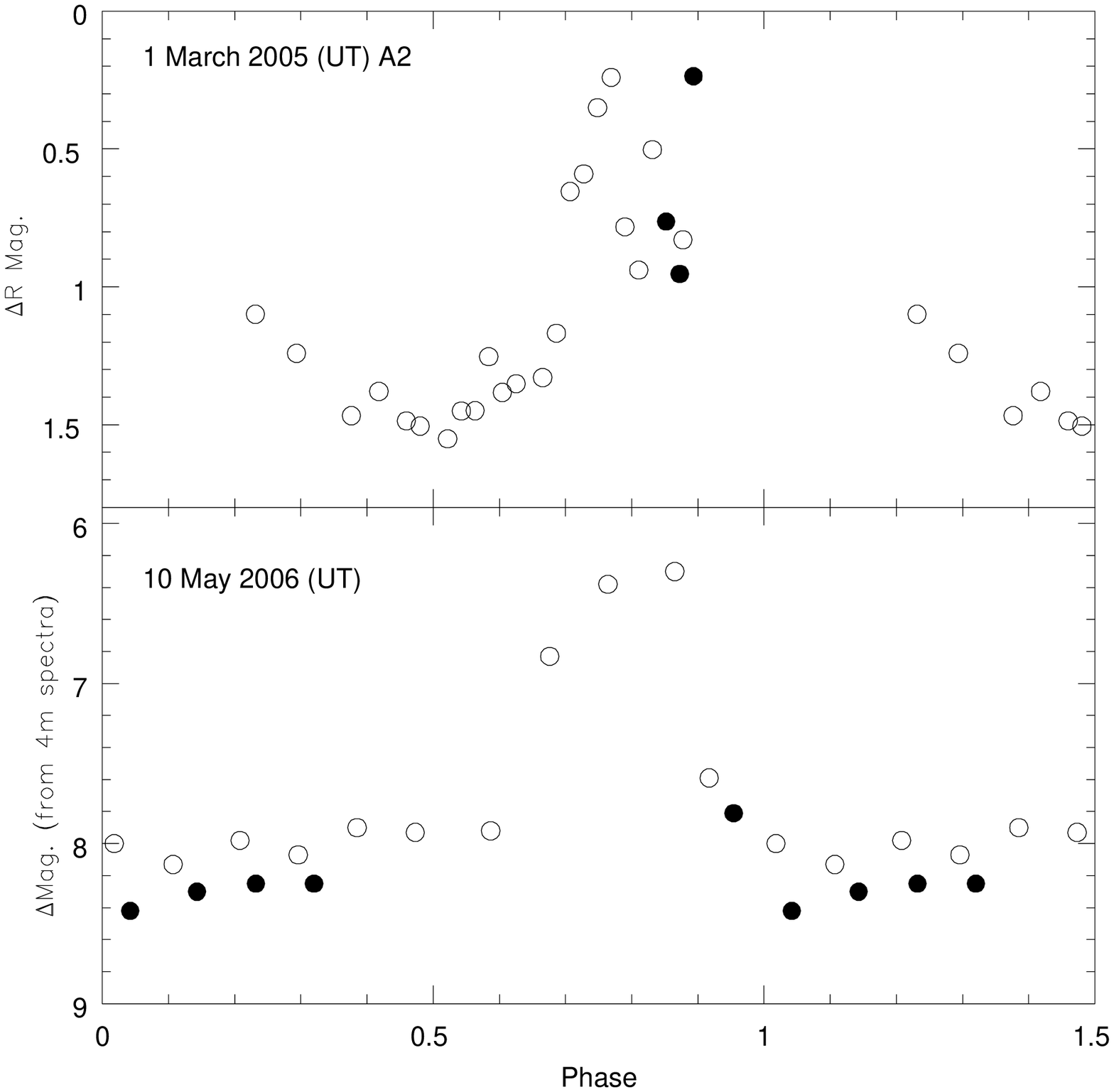}
\caption{Left: Phased light curve of ST LMi from observing run D, as marked in figure 1 
(bottom).
Note the weakness of the maximum near phase 0.8 compared to figure 2 (left), and the
near absence of scatter in the light curve compared to figure 2 and figure 4 (bottom).
Filled symbols are data from the second of two adjacent orbital cycles.  
Right: Phased R-band light 
curves of ST LMi.  Top is a differential R-band light curve from
run A2, as marked in figure 1 (bottom).  Bottom is a synthetic light curve extracted from
the 4-m spectra of 2006-May-19, with an effective wavelength approximately midway
between V and R.  Note the large amplitude of the maximum near phase 0.8 in both plots.}
\end{figure}

\begin{figure}  
\epsscale{1.0}
\plotone{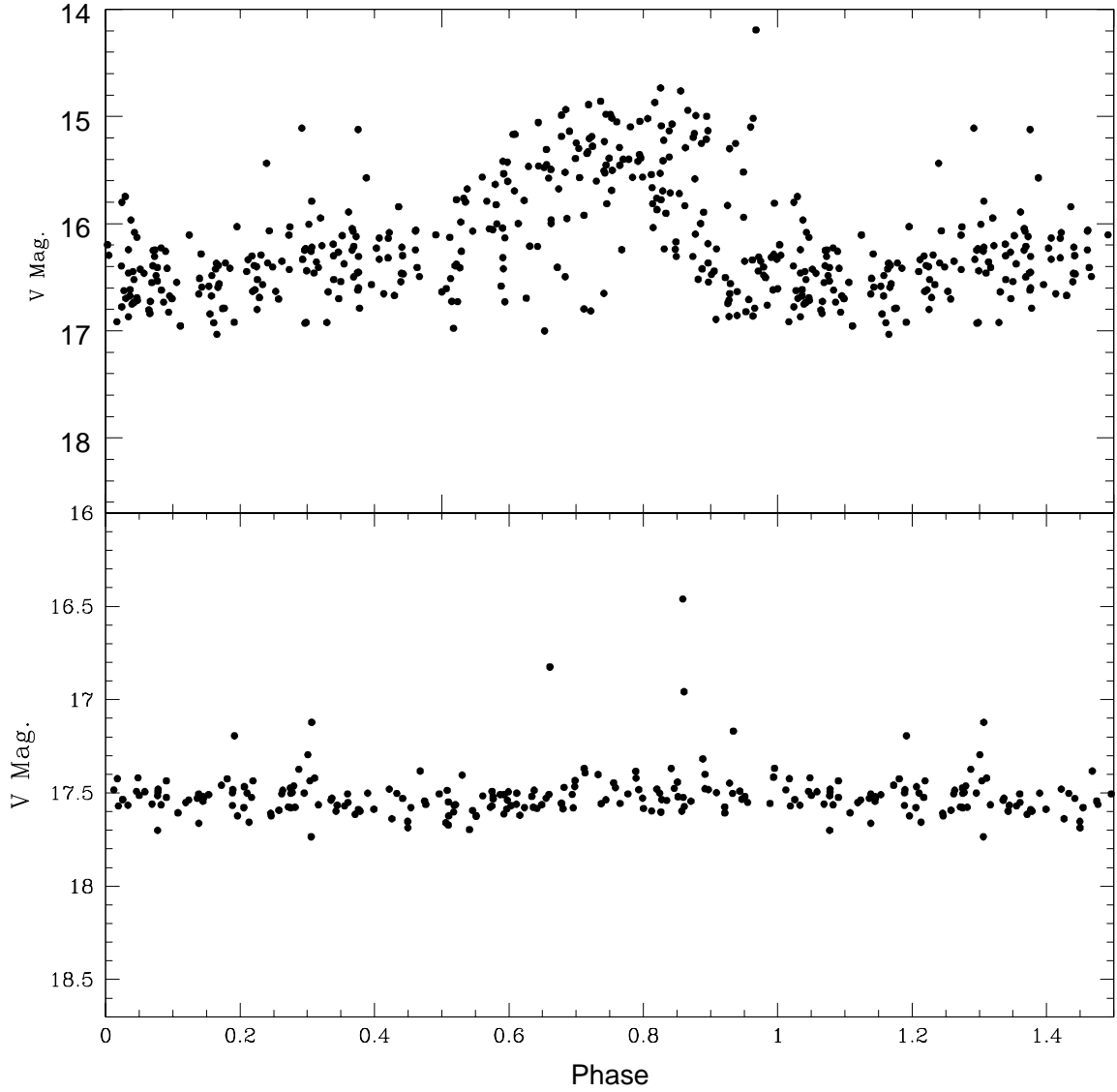}
\caption{Phased V-band light curves of ST LMi from RoboScope data 1990-2005.  The data 
consist of
single point each clear night, at a random orbital phase.  Top is the high state light
curve showing a maximum near phase 0.8.  Bottom is the low-state light curve, perhaps
showing a weak maximum near phase 0.8.  The occasional brightward excursions in the bottom panel
are likely artifacts due to cosmic rays.}
\end{figure}

\begin{figure}  
\epsscale{1.0}
\plotone{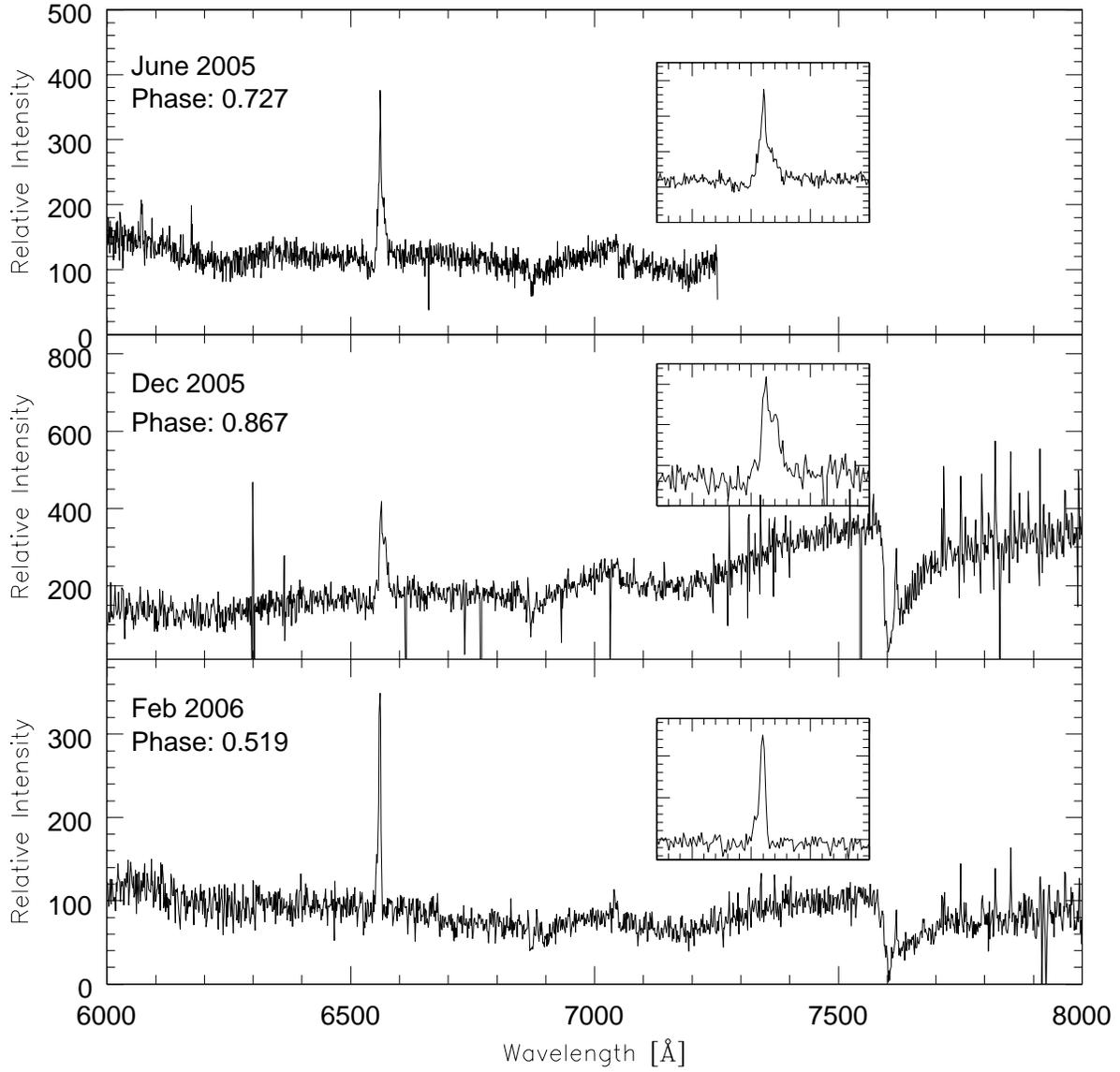}
\caption{Representative low-state spectra of ST LMi from 2005-Jun (top), 2005-Dec 
(middle) and
2006-Feb (bottom).  No corrections for atmospheric absorption have been made.  The
middle panel shows sharp features due to incomplete cancellation of night
sky emission lines.  The inset plots show the H$\alpha$ emission line on an expanded
scale.}
\end{figure}

\begin{figure}  
\epsscale{1.0}
\plotone{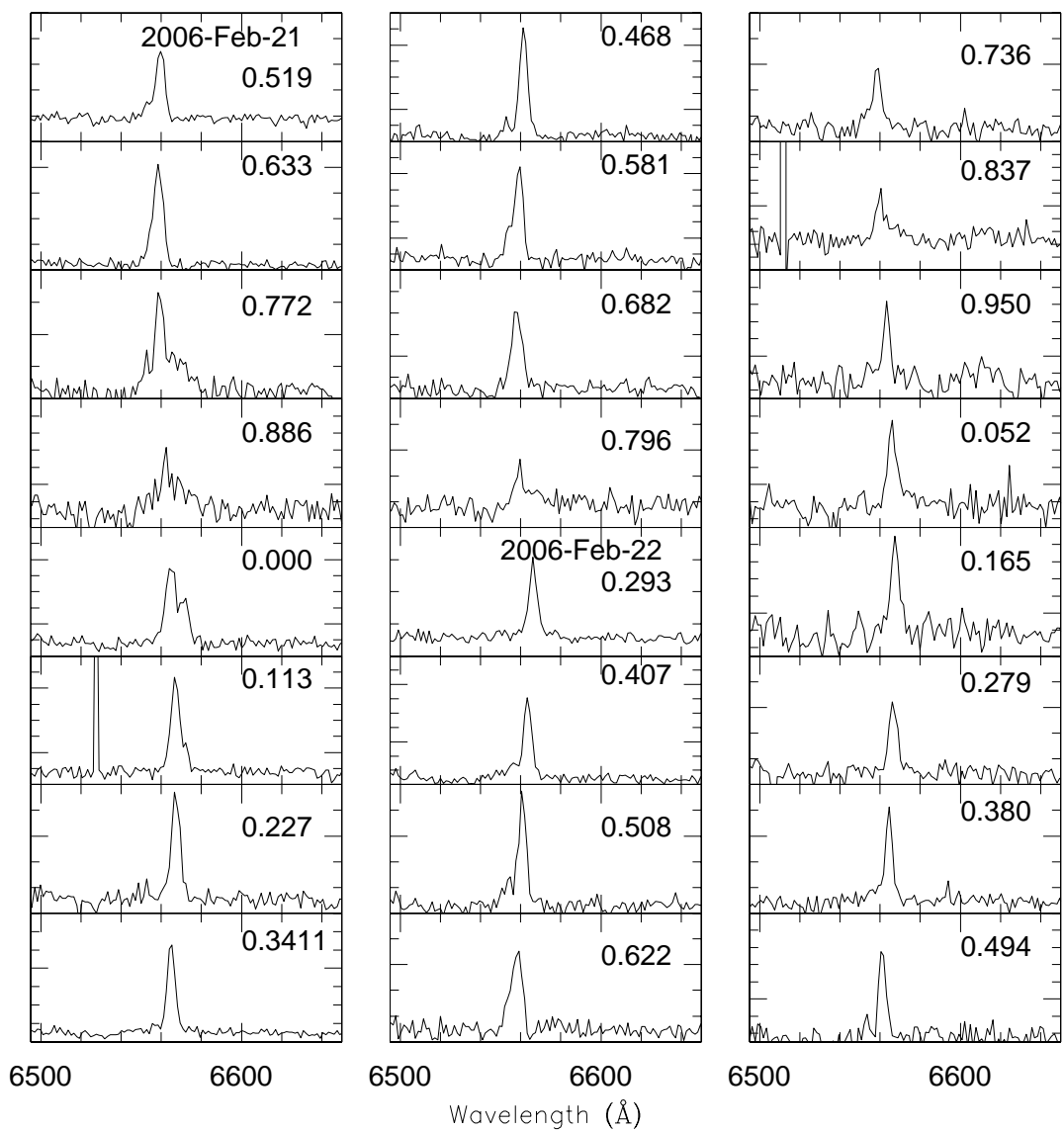}
\caption{Low-state H$\alpha$ profiles of ST LMi for orbital phases 0.5192 to 0.7963 on
2006-Feb-21 UT (first night) and for phases 0.2930 to 0.4942 on 2006-Feb-22 UT (second night)).  
Successive phases run down the page, starting in the upper left corner.}
\end{figure}

\begin{figure}  
\epsscale{1.0}
\plotone{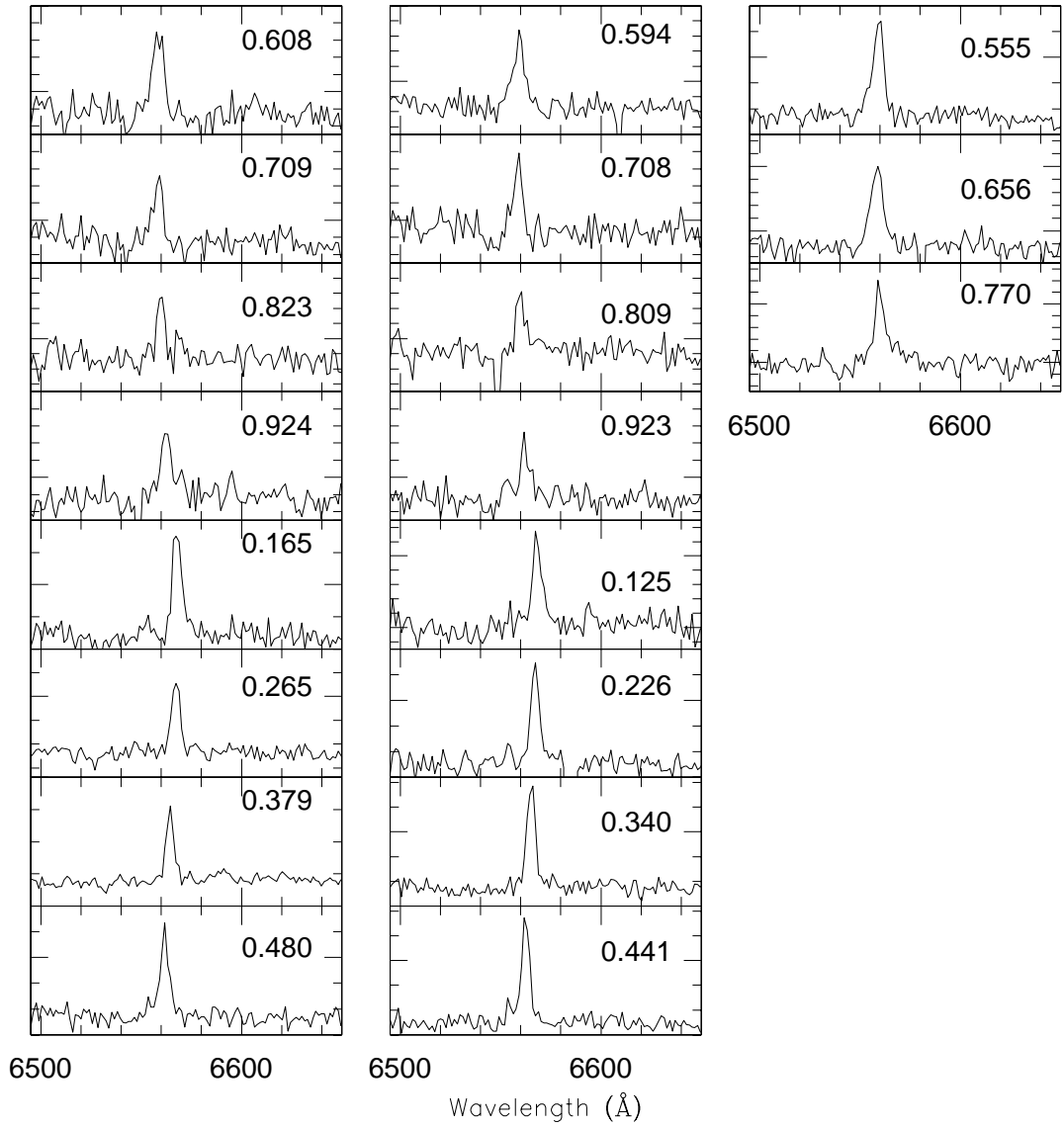}
\caption{Continuation of previous figure (for night 2).}
\end{figure}

\begin{figure}  
\epsscale{1.0}
\plotone{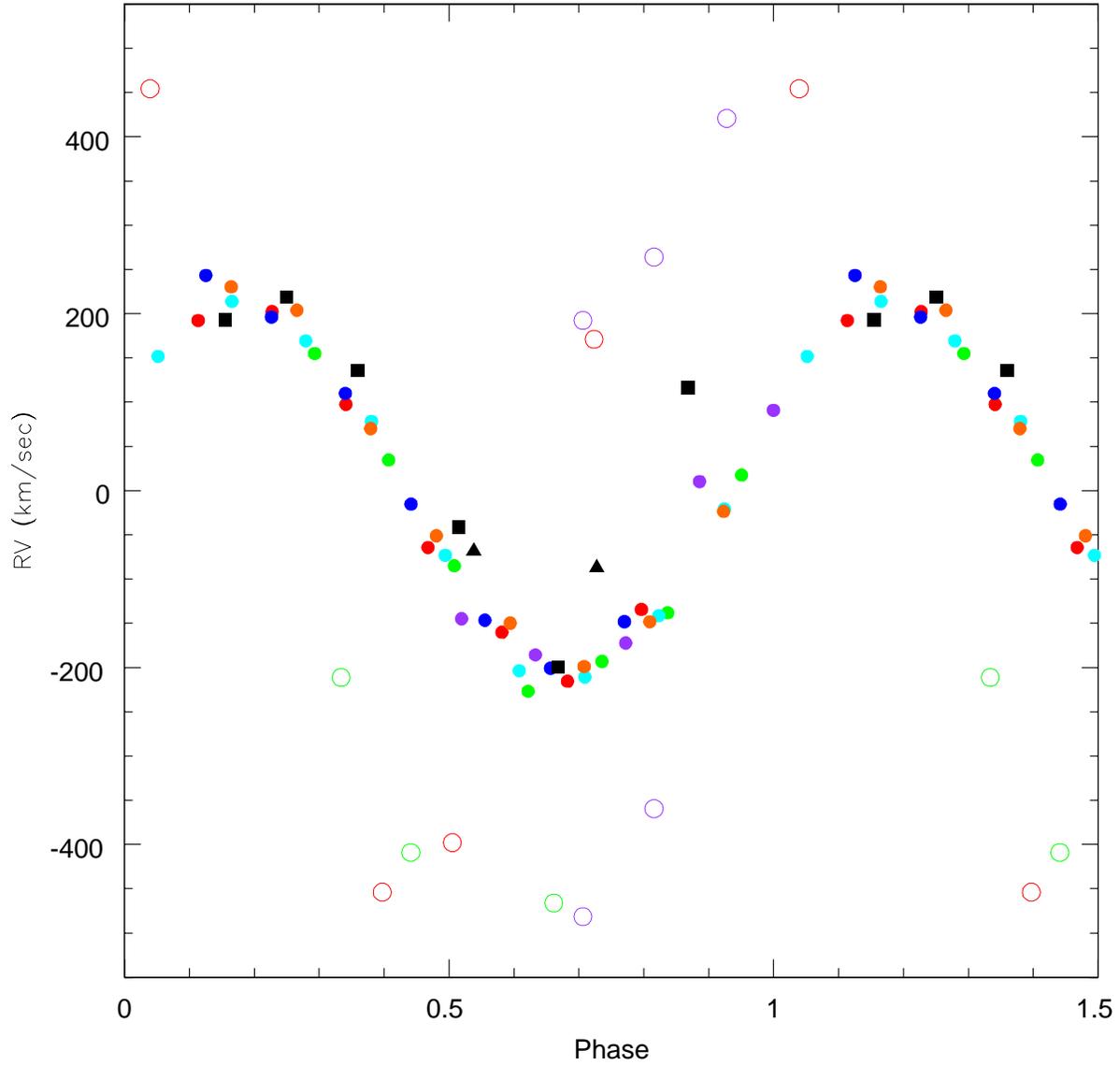}
\caption{The RVs of the central H$\alpha$ emission line component in ST LMi for the low-state 
WIYN spectra of 2005-June-17 (squares), the 4-m spectra of 2005-Dec-31 (triangles), and 
the low-state WIYN spectra of  2006-Feb-21/22 (solid points). Open circles represent the 
RVs  of the  occasional satellite components in the WIYN 2006-Feb spectra. Colors represent 
points of the same orbit.}
\end{figure}  

\begin{figure}  
\epsscale{1.0}
\plotone{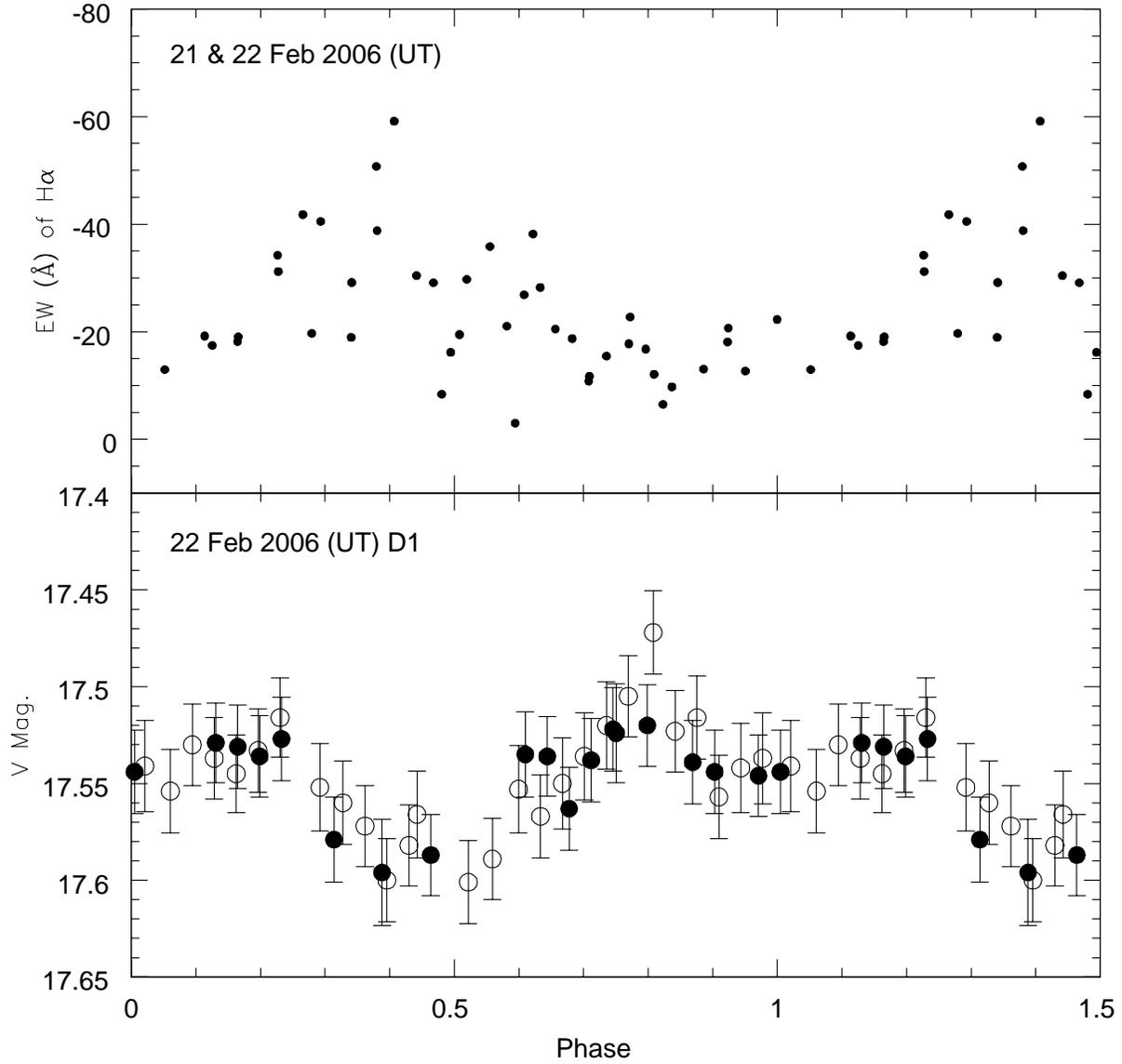}
\caption{Top: Total EW of the low-state WIYN H$\alpha$ emission line vs orbital phase; 
Bottom: The concurrent phase light curve.}
\end{figure}

\begin{figure}  
\epsscale{1.0}
\plotone{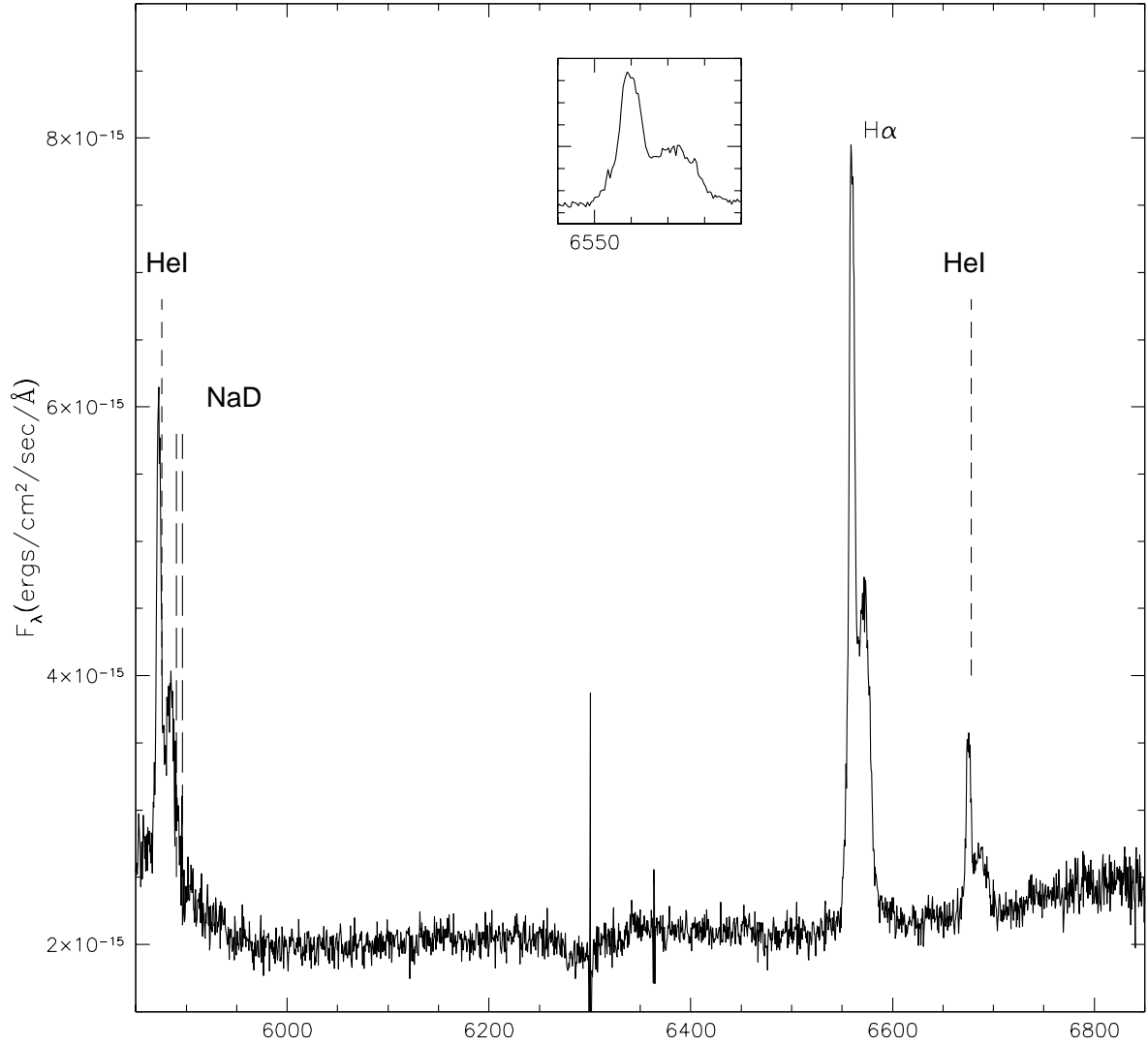}
\caption{A representative high-state spectrum of ST LMi from the 2006-May-19 (UT) 4-m run.  
The inset plot is an expanded view of the H$\alpha$ profile.}
\end{figure}

\begin{figure}  
\epsscale{1.1}
\plottwo{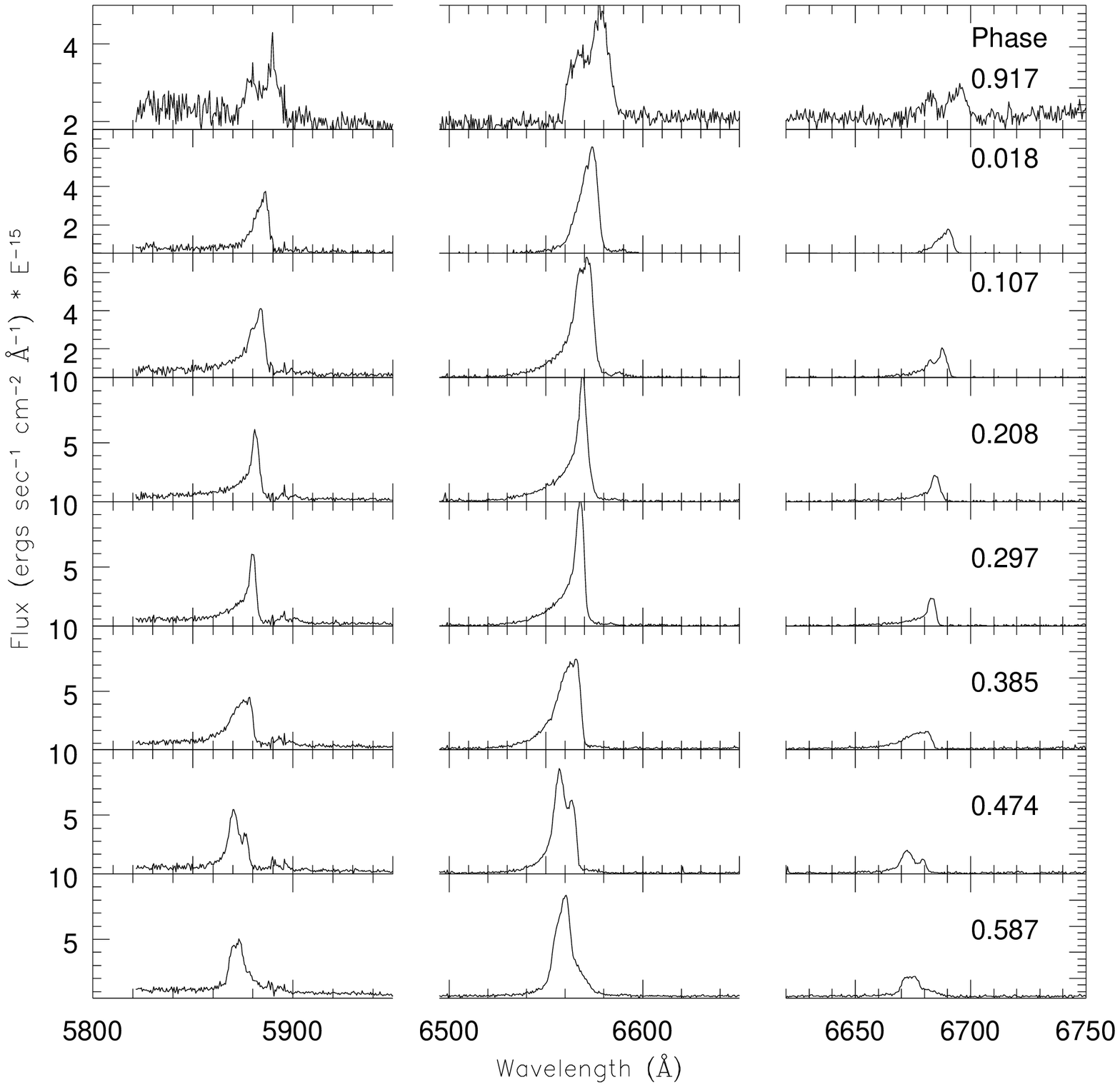}{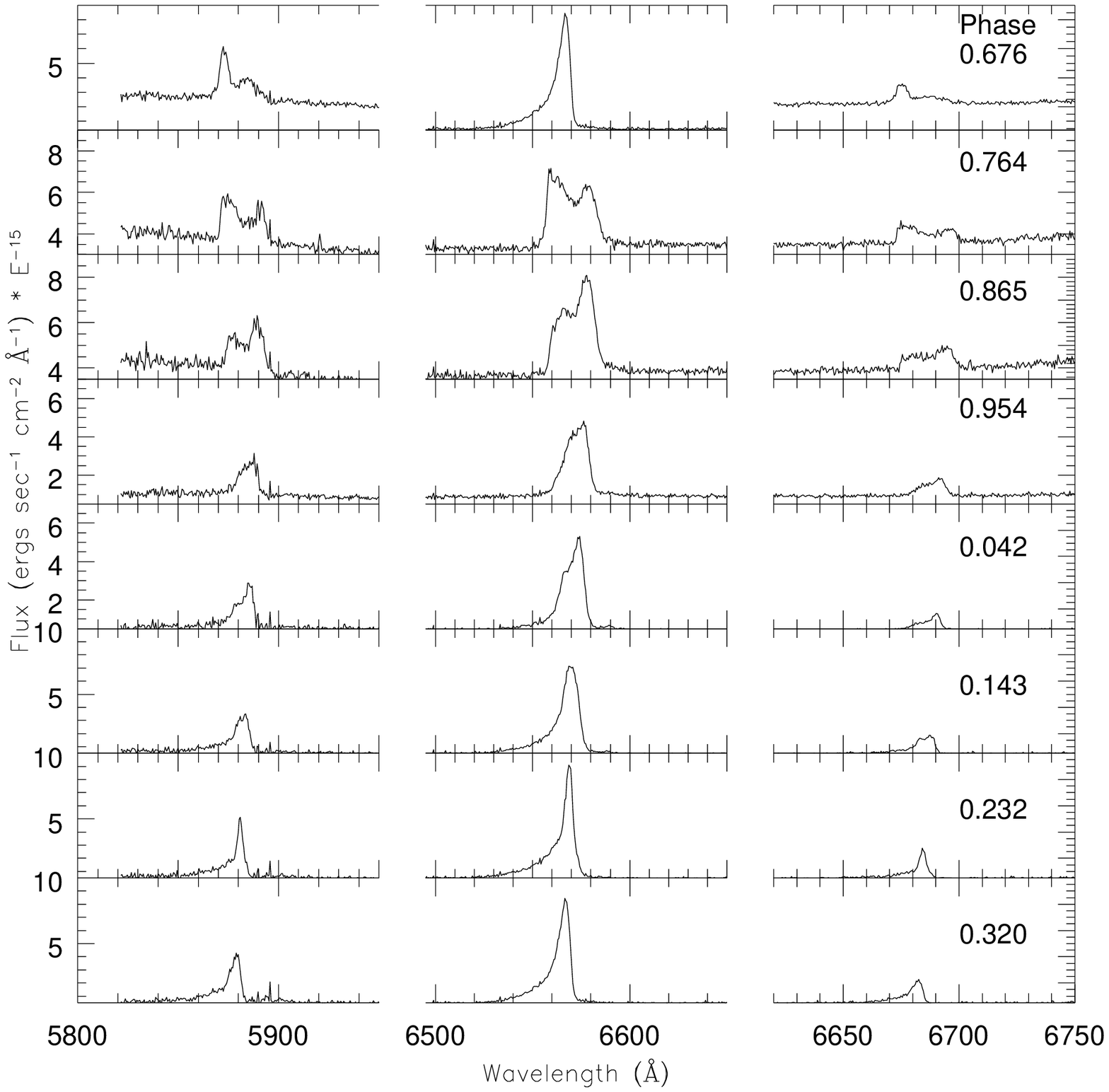}
\caption{Line profiles of HeI 5876$\AA$ (left column), H$\alpha$ (center column) and 
HeI 6678$\AA$ (right column) as a function of orbital phase, for the high-state spectra
of 2006-May-19 2006 UT.  The NaD emission lines (from Tucson airglow) are often visible near and sometimes within the red wing of HeI 5876$\AA$.}
\end{figure}

\begin{figure}  
\epsscale{1.0}
\plotone{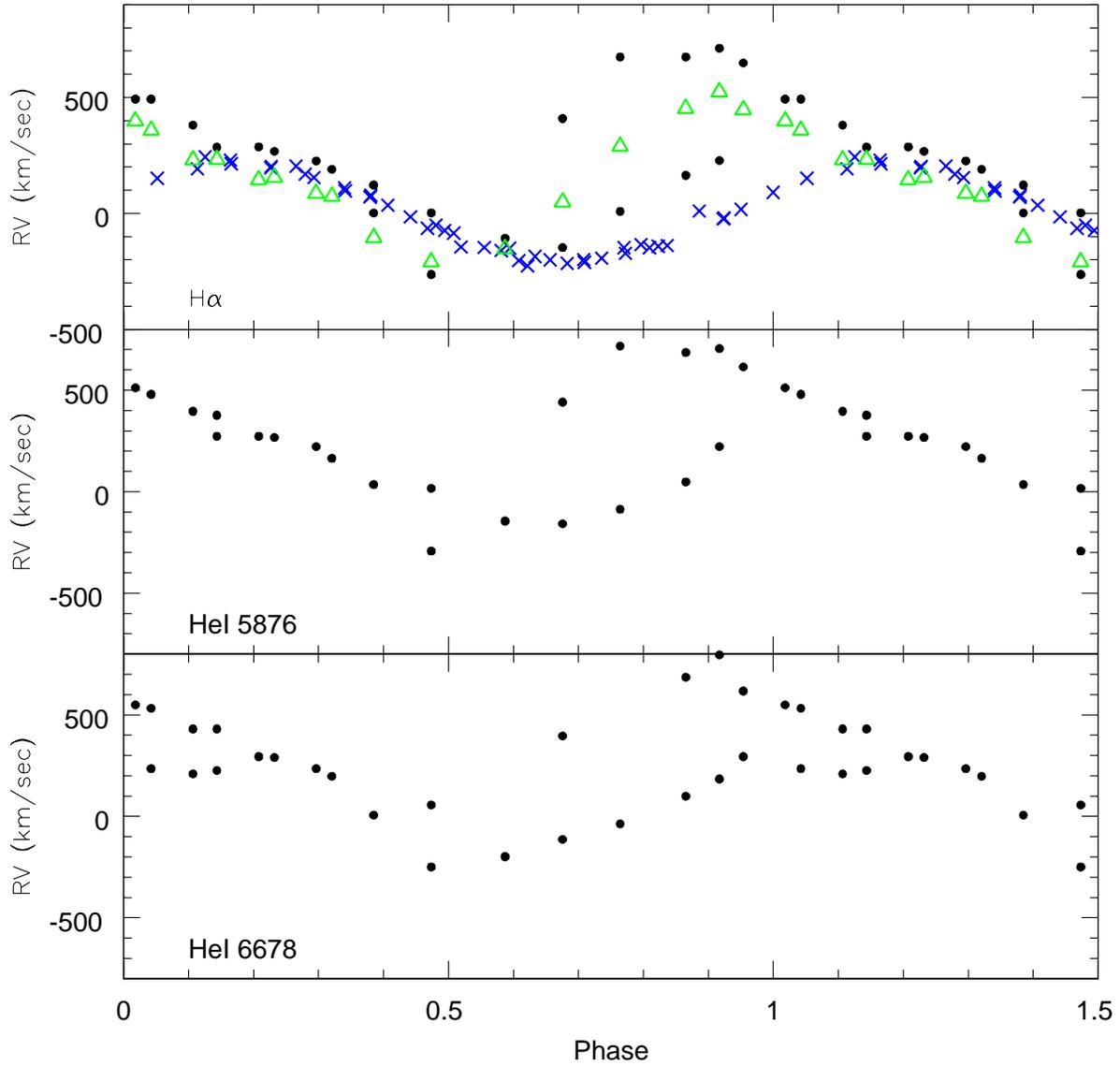}
\caption{High-state (2006-May-19 UT) radial velocity curves of ST LMi for H$\alpha$ (top), 
and two HeI emission lines (middle and bottom).  The open triangles are velocities of the
centroid of H$\alpha$.  The filled circles are velocities at the peaks of the most 
distinct one or two  emission line components, as seen in the line profiles in figure 10.  For 
comparison, the crosses in the top panel are the ST LMi H$\alpha$ velocities for the 
low-state WIYN data of 2006-Feb-21/22 UT.}
\end{figure}

\begin{figure}  
\epsscale{1.1}
\plotone{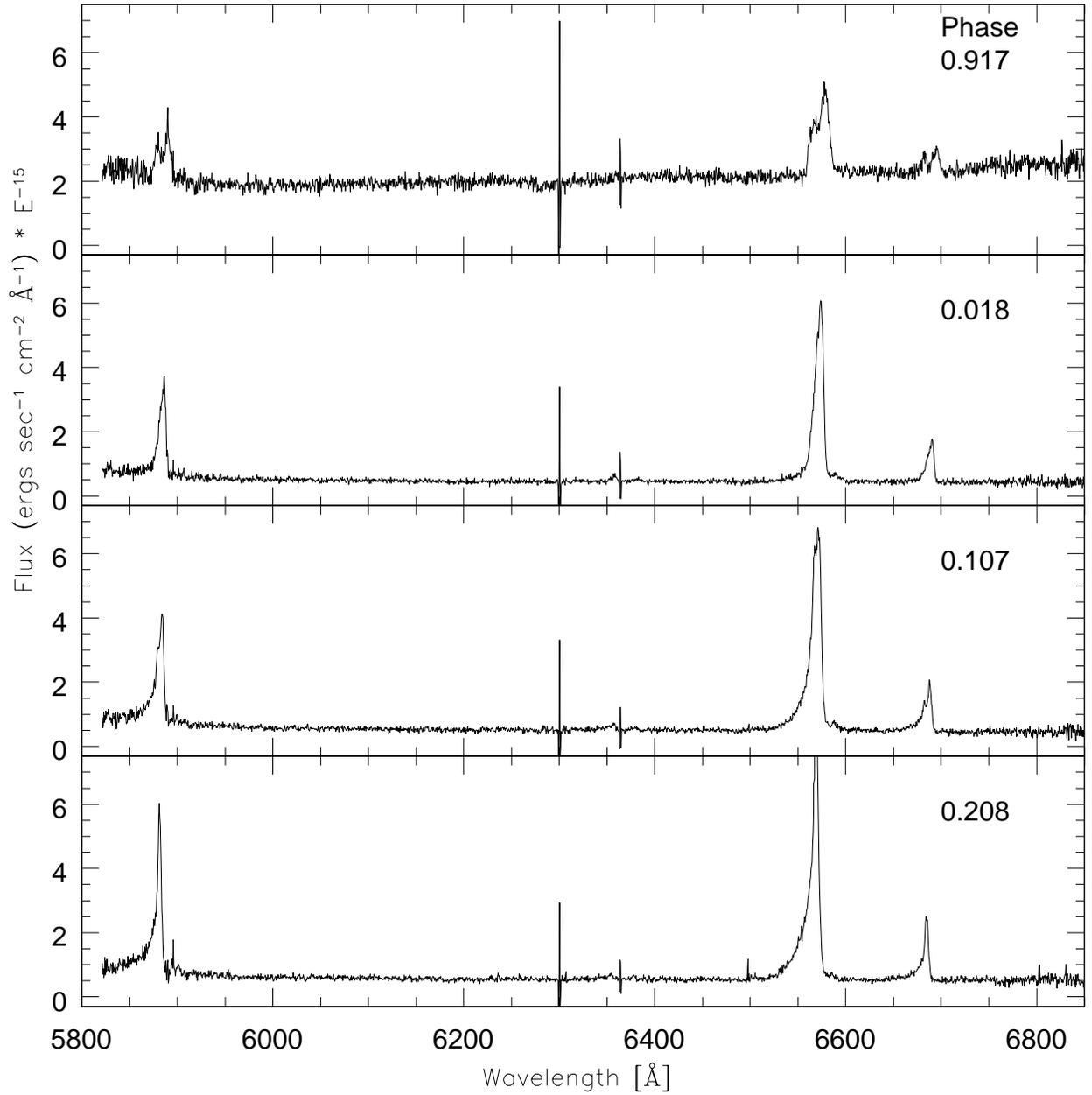}
\caption{The complete spectral region of ST LMi as a function of orbital phase
for the high-state 2006-May-19 (UT) data. We are using the same scale for all spectra, to emphasize the continuum and the weaker
spectral features. (See text for discussion.)}
\end{figure}

\begin{figure}  
\epsscale{1.1}
\plotone{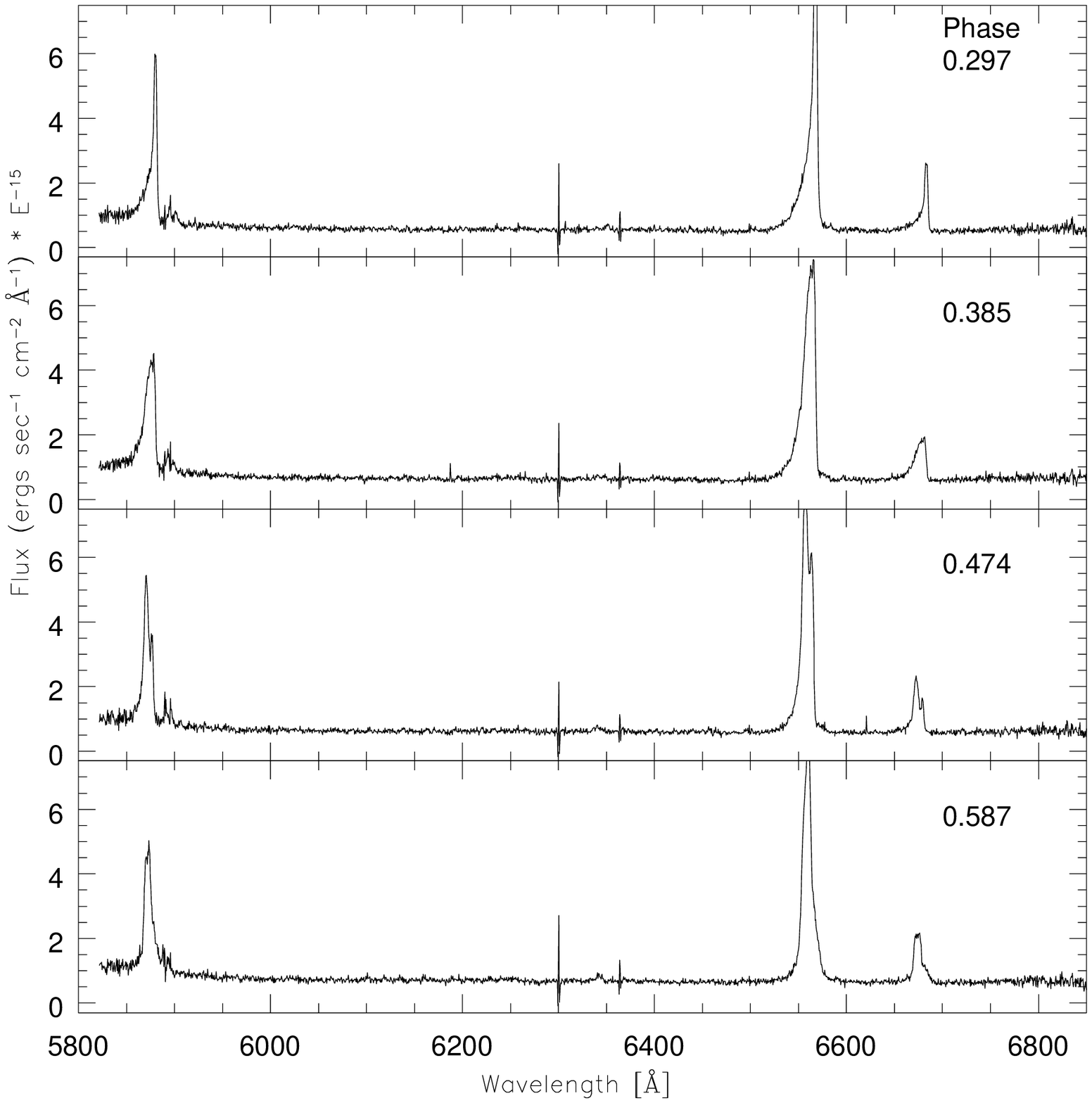}
\caption{The complete spectral region of ST LMi as a function of orbital phase
for the high-state 2006-May-19 (UT) data (cont).}
\end{figure}

\begin{figure}  
\epsscale{1.1}
\plotone{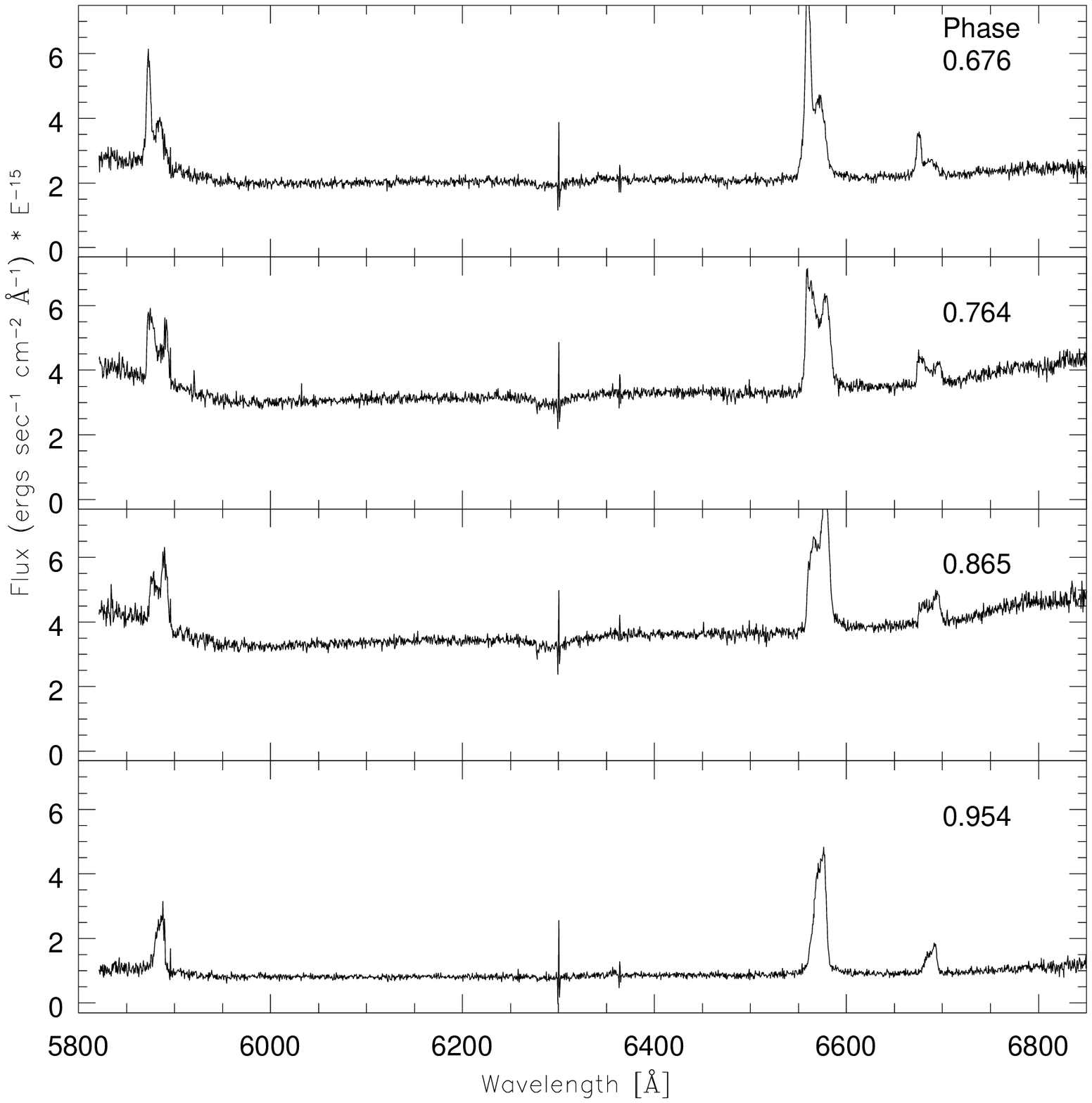}
\caption{The complete spectral region of ST LMi as a function of orbital phase
for the high-state 2006-May-19 (UT) data (cont)}
\end{figure}

\begin{figure}  
\epsscale{1.1}
\plotone{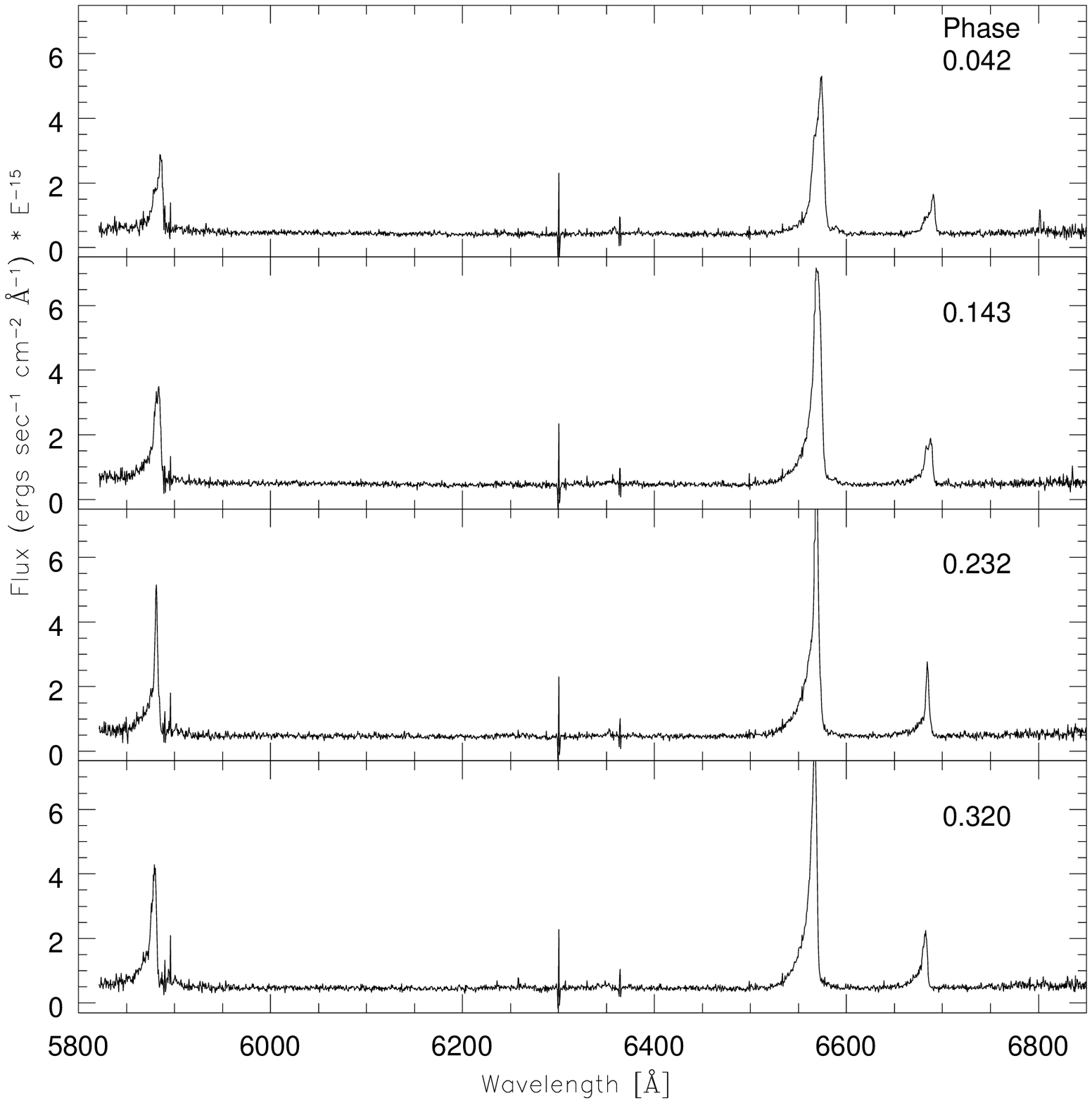}
\caption{The complete spectral region of ST LMi as a function of orbital phase
for the high-state 2006-May-19 (UT) data (cont)}
\end{figure}

\begin{figure}  
\epsscale{1.5}
\plottwo{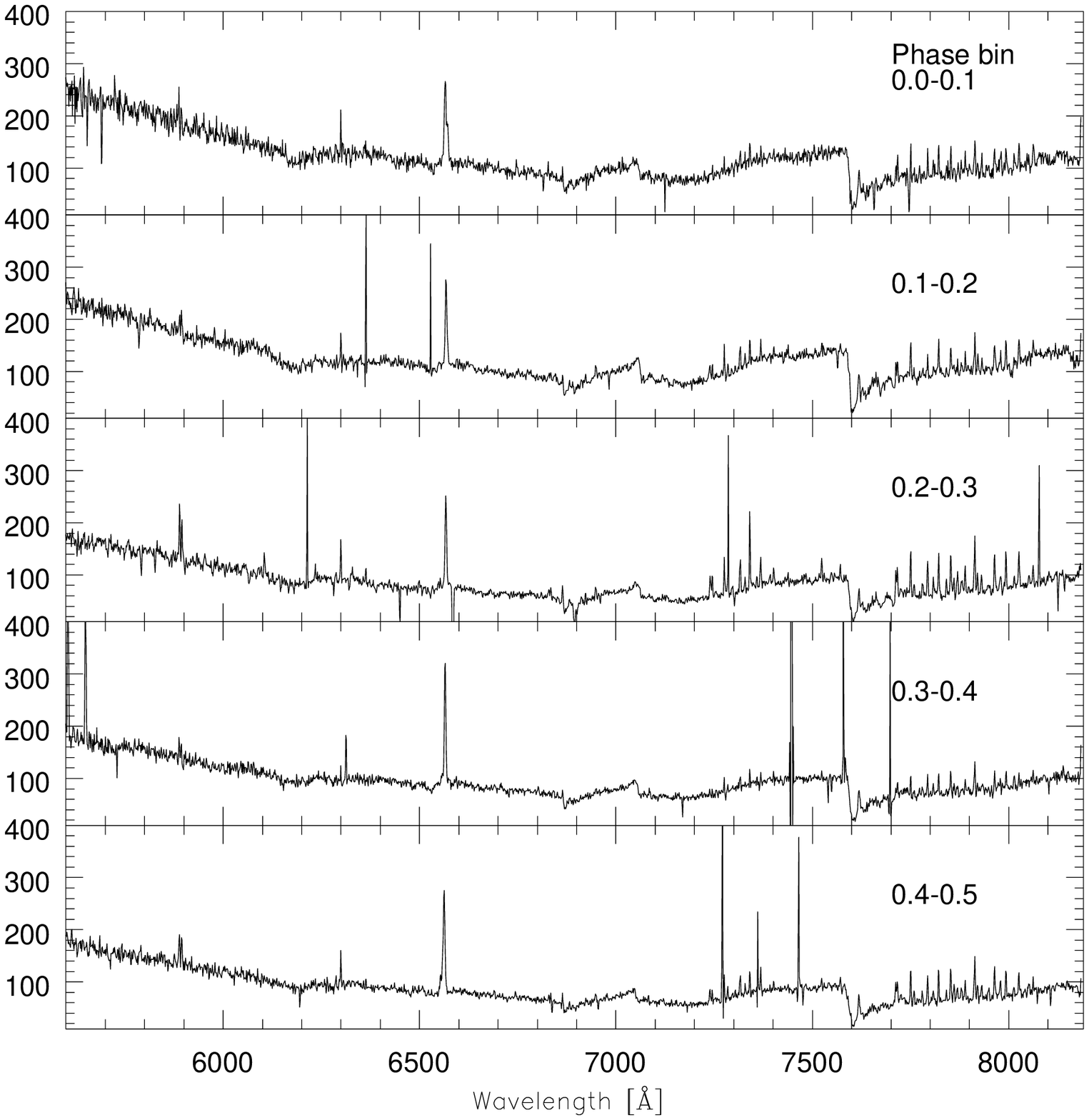}{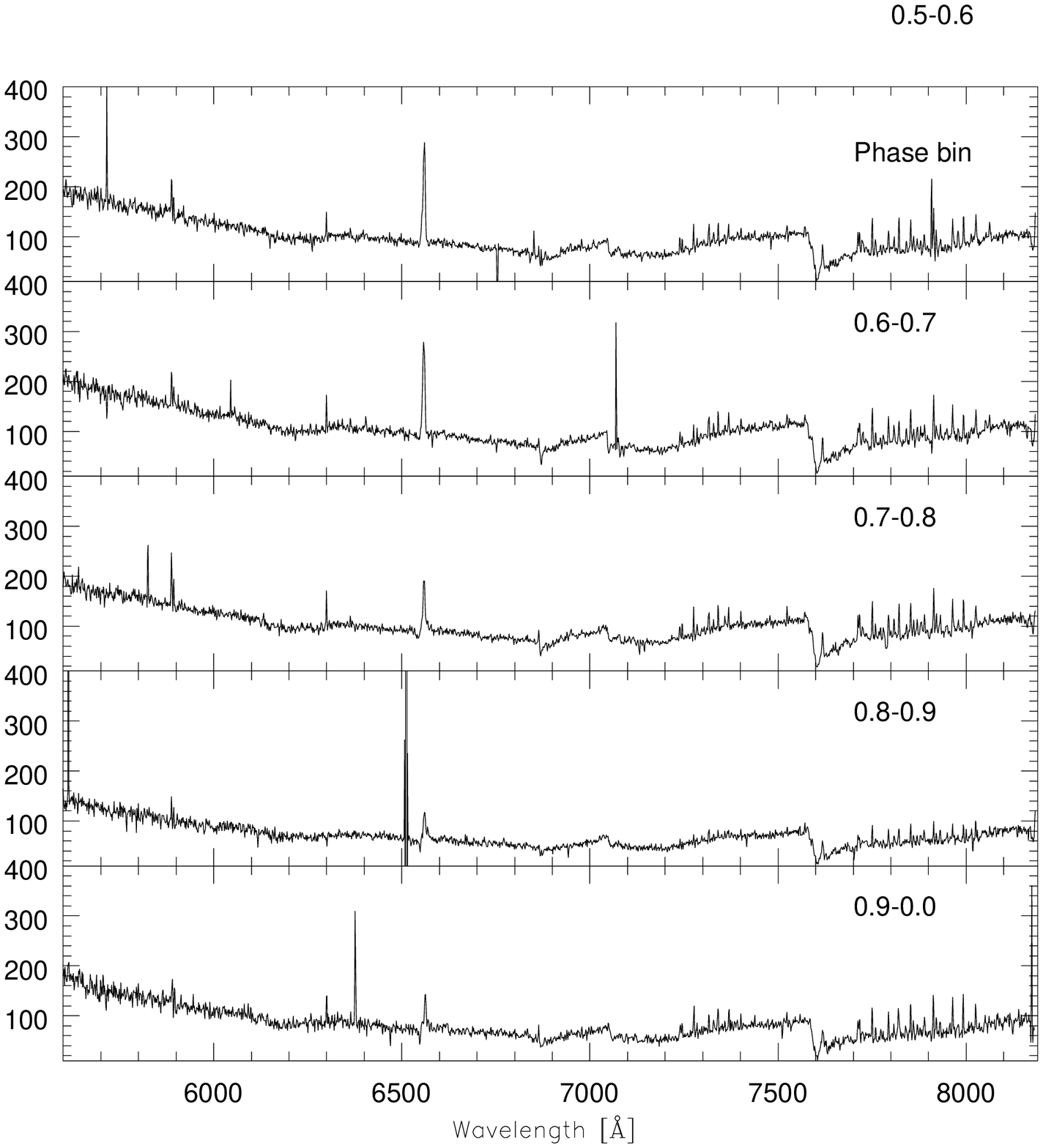}
\caption{WIYN low-state spectra from 2006-Feb-21/22 UT, averaged in phase bins of 0.1. The narrow weak emission lines, mostly redward of 7200$\AA$, are
artifacts due to incomplete cancellation of OH features in the
night sky from the fiber spectra. (Also see text for comments.)}
\end{figure}

\end{document}